\documentclass{aa}  

\usepackage{graphicx}   % Including figure files
\usepackage{txfonts}
\usepackage[colorlinks=true,linkcolor=blue,citecolor=blue,filecolor=black,urlcolor=blue]{hyperref}

\begin{document} 

   \title{AT 2019avd: A novel addition to the diverse population of nuclear transients}
   \author{A. Malyali\inst{1}
          \and
A.~Rau\inst{1} \and
A.~Merloni\inst{1} \and
K.~Nandra\inst{1} \and
J.~Buchner\inst{1} \and
Z.~Liu\inst{1} \and
S.~Gezari\inst{2,3} \and
J.~Sollerman\inst{4} \and
B.~Shappee\inst{5} \and
B.~Trakhtenbrot\inst{6} \and
I.~Arcavi\inst{6,7} \and
C.~Ricci\inst{8,9} \and
S.~van Velzen\inst{10,2,11} \and
A.~Goobar\inst{12} \and
S.~Frederick\inst{2} \and
A.~Kawka\inst{13} \and
L.~Tartaglia\inst{4,14} \and
J.~Burke\inst{15,16} \and 
D.~Hiramatsu\inst{15,16} \and
M.~Schramm\inst{17} \and
D.~van der Boom\inst{6} \and
G.~Anderson\inst{13} \and
J.~C.~A.~Miller-Jones\inst{13} \and
E.~Bellm\inst{18} \and
A.~Drake\inst{19} \and
D.~Duev\inst{19} \and
C. Fremling\inst{19} \and
M.~Graham\inst{19} \and
F.~Masci\inst{20} \and
B.~Rusholme\inst{20} \and
M.~Soumagnac\inst{21,22} \and
R.~Walters\inst{23}
          }
   \institute{Max-Planck-Institut f\"ur extraterrestrische Physik,  Giessenbachstrasse 1, 85748 Garching, Germany\\
              \email{amalyali@mpe.mpg.de}
         \and
Department of Astronomy, University of Maryland, College Park, MD 20742, USA
         \and
Space Telescope Science Institute, Baltimore, MD 21218, USA
         \and
Department of Astronomy and the Oskar Klein Centre, Stockholm University, AlbaNova, SE 10691 Stockholm, Sweden
         \and
Institute for Astronomy, University of Hawaii at Manoa, 2680 Woodlawn Dr., Honolulu, HI 96822
         \and
The School of Physics and Astronomy, Tel Aviv University, Tel Aviv 69978, Israel
         \and
CIFAR Azrieli Global Scholars program, CIFAR, Toronto, Canada
         \and
N\'ucleo de Astronom\'ia de la Facultad de Ingenier\'ia, Universidad Diego Portales, Av. Ej\'ercito Libertador 441, Santiago, Chile
         \and
Kavli Institute for Astronomy and Astrophysics, Peking University, Beijing 100871, China
         \and
Center for Cosmology and Particle Physics, New York University, NY 10003, USA
         \and
Leiden Observatory, Leiden University, PO Box 9513, 2300 RA Leiden, The Netherlands
        \and
The Oskar Klein Centre, Department of Physics, AlbaNova, Stockholm University, SE 10691 Stockholm, Sweden
         \and
International Centre for Radio Astronomy Research - Curtin University, GPO Box U1987, Perth, WA 6845, Australia
         \and
INAF - Osservatorio Astronomico di Trieste, Via G.B.~Tiepolo, 11, I-34143 Trieste, Italy
         \and
Department of Physics, University of California, Santa Barbara, CA 93106-9530, USA
         \and
Las Cumbres Observatory, 6740 Cortona Dr, Suite 102, Goleta, CA 93117-5575, USA
         \and
Graduate School of Science and Engineering, Saitama Univ., 255 Shimo-Okubo, Sakura-ku, Saitama City, Saitama 338-8570, Japan
        \and
DIRAC Institute, Department of Astronomy, University of Washington, 3910 15th Avenue NE, Seattle, WA 98195, USA
        \and 
Division of Physics, Mathematics, and Astronomy, California Institute of Technology, Pasadena, CA 91125, USA
        \and 
IPAC, California Institute of Technology, 1200 E. California Boulevard, Pasadena, CA 91125, USA
        \and
Lawrence Berkeley National Laboratory, 1 Cyclotron Road, Berkeley, CA 94720, USA
        \and
Department of Particle Physics and Astrophysics, Weizmann Institute of Science, Rehovot 76100, Israel
        \and
California Institute of Technology, Pasadena, CA 91125, USA
             }

   \date{Received XXX; accepted YYY}

\abstract{We report on {\it SRG}/eROSITA, ZTF, ASAS-SN, Las Cumbres, NEOWISE-R, and {\it Swift} XRT/UVOT observations of the unique ongoing event AT~2019avd, located in the nucleus of a previously inactive galaxy at $z=0.029$. eROSITA first observed AT~2019avd on 2020-04-28 during its first all sky survey, when it was detected as an ultra-soft X-ray source ($kT\sim 85$~eV) that was $\gtrsim 90$ times brighter in the $0.2-2$~keV band than a previous 3$\sigma$ upper flux detection limit (with no archival X-ray detection at this position). The ZTF optical light curve in the $\sim 450$ days preceding the eROSITA detection is double peaked, and the eROSITA detection coincides with the rise of the second peak. Follow-up optical spectroscopy shows the emergence of a Bowen fluorescence feature and high-ionisation coronal lines ([\ion{Fe}{X}] 6375~{\AA}, [\ion{Fe}{XIV}] 5303~{\AA}), along with persistent broad Balmer emission lines (FWHM$\sim 1400$~km~s$^{-1}$). Whilst the X-ray properties make AT~2019avd a promising tidal disruption event (TDE) candidate, the optical properties are atypical for optically selected TDEs. We discuss potential alternative origins that could explain the observed properties of AT~2019avd, such as a stellar binary TDE candidate, or a TDE involving a super massive black hole binary.}


   \maketitle
%
%-------------------------------------------------------------------

\section{Introduction}

Actively accreting supermassive black holes (SMBHs) have long been known to exhibit large amplitude flaring behaviour (e.g. \citealt{Tohline1976,ANTONUCCI1983,Penston1984,Shappee2014,Storchi-Bergmann2017,Frederick2019}), whereby multi-epoch observations of galaxy nuclei, over year-long timescales, have revealed drastic changes in their luminosity. The physical mechanisms responsible for producing extreme accretion rate changes are still unclear, although various models have been suggested, such as state transitions in the inner disc \citep{Noda2018,Ross2018}, radiation pressure instabilities in the disc \citep{Sniegowska2019}, or tidal disruption events (TDEs; \citealt{Merloni2015,Chan2019}).

Whilst the sample of ignition events in galactic nuclei was previously limited to only a few objects, the advance of wide-field, high-cadence surveys over the last decade has facilitated the discovery of an increasing number of extreme state changes. This has resulted in tighter constraints on the timescales of flaring events for these systems. For example, \citet{Trakhtenbrot2019a} recently reported a new class of SMBH accretion event that sees a large amplitude rise in the optical/UV luminosity over timescales of months. 

In addition to triggering drastic changes in the accretion rate in AGNs, TDEs can also cause quiescent black holes to transition into short-lived active phases. In a TDE, a star that passes too close to a BH is torn apart by strong tidal forces, with a fraction of the bound stellar debris then being accreted onto the BH \citep{Hills1975,Young1977,Gurzadian1981,Lacy1982,Rees1988,Phinney1988}. Early TDE candidates were first identified through detection of large-amplitude (at least a factor of 20), ultra-soft X-ray flares (black-body temperatures between 40 and 100 eV) from quiescent galaxies during the ROSAT survey \citep{Bade1996,Komossa1999a,Komossa1999,Grupe1999,Greiner2000}. Since then, the vast majority of TDE candidates have been optically selected, such as through the Sloan Digital Sky Survey (SDSS; e.g. \citealt{VanVelzen2011a,Merloni2015}), the Panoramic Survey Telescope and Rapid Response System (Pan-STARRS; e.g. \citealt{Gezari2012,Holoien2019}), the Palomar Transient Factory (PTF; e.g. \citealt{Arcavi2014}), the Intermediate Palomar Transient Factory (iPTF;  e.g. \citealt{Blagorodnova2017a,Hung2017}), the All Sky Automated Survey for SuperNovae (ASAS-SN; e.g. \citealt{Holoien2014, Holoien2016,Wevers2019a,Holoien2019a}), and the Zwicky Transient Facility (ZTF;  e.g. \citealt{VanVelzen2019,van2020seventeen}). Optically selected TDEs are characterised as blue nuclear transients with light curves showing longer/ shorter rise and decay timescales relative to supernovae (SNe)/ AGN\footnote{For large, well-defined AGN flares similar to those seen in \citet{Frederick2019}, as opposed to stochastic AGN variability.}, and a relatively smooth power-law decline. Optical spectroscopic follow-up of these events post-peak reveals blue continua (blackbody temperatures $\sim 10^4$K) with various broad emission lines (full width at half maximum, FWHM $\lesssim 10^4$~km~s$^{-1}$); a recent characterisation of the different TDE spectroscopic classes was presented by \citet{van2020seventeen}.
Although a number of TDE candidates have also been found through UV selection ({\it GALEX}, \citealt{Gezari2008,Gezari2009}), and X-ray selection ({\it XMM-Newton} Slew, \citealt{Esquej2007,Esquej2008b,Saxton2012a,Saxton2017}), most of our understanding of TDEs is currently biased towards this set of observed properties of optically-selected TDEs.

Whilst most previous TDE searches focused on identifying TDEs in quiescent galaxies, an increasing number of candidates for TDEs in AGNs are being proposed in the literature \citep{Merloni2015,Blanchard2017PS16dtm:Galaxy,Trakhtenbrot2019,Liu2020,Ricci2020}. In certain cases, the distinction between TDE and non-TDE-induced SMBH accretion state changes is becoming increasingly blurred (see also \citealt{Neustadt2020}). Variants of TDEs have also been proposed to explain more exotic phenomena, such as the recently observed quasi-periodic eruptions (QPEs) in a few galactic nuclei \citep{Miniutti2019Nine-hourNucleus,Giustini2020,King2020}, and periodic flaring seen in an AGN \citep{Payne2020}. Other origins for extreme nuclear transients involve SNe in the AGN accretion disc \citep{Rozyczka1995}, or interaction of SMBH binaries (SMBHB) with an accretion disc \citep{Kim2018}. It is clear that such different physical origins may result in a diverse range of observed variability behaviours.

\begin{figure}
    \centering
    \includegraphics[scale=1.2,trim={0.25cm 0.3cm 0 0},clip]{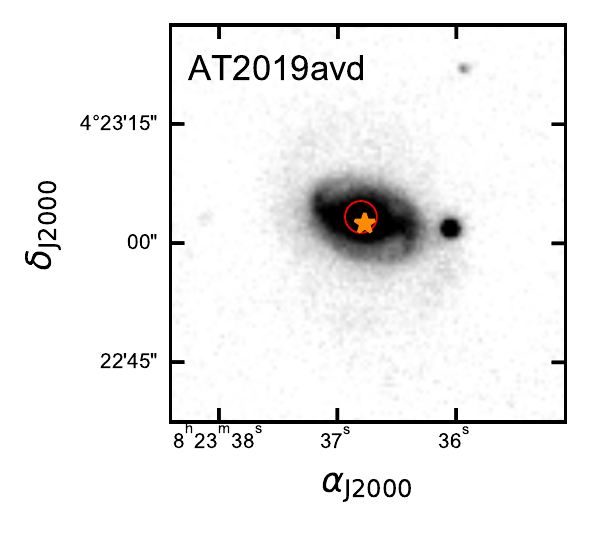}
    \caption{Pan-STARRS $g$-band image centred on the host galaxy of AT~2019avd. The dark orange star and red circle mark the ZTF position and eROSITA localisation respectively, where the radius of the circle is set to the 2$^{\prime\prime}$ uncertainty on the eROSITA source position.}
    \label{fig:finder_chart}
\end{figure}

 In this paper, we report on the ongoing extreme event AT~2019avd, which is a novel addition to the already diverse population of nuclear transients. 
 AT~2019avd is associated to the previously inactive galaxy 2MASX~J08233674+0423027 at $z=0.029$ (see Fig.~\ref{fig:finder_chart}), and was first reported as ZTF19aaiqmgl at the Transient Name Server (TNS\footnote{\url{https://wis-tns.weizmann.ac.il/}}) following its discovery by ZTF on 2019-02-09 UT\footnote{all dates in this paper will be reported in UT format.} \citep{2019TNSTR.236....1N}. The transient was independently detected more than a year later on 2020-04-28 as a new ultra-soft nuclear X-ray source \citep{2020ATel13712....1M} during the first all-sky survey of the eROSITA instrument (Predehl et al., in press) on-board the Russian/German Spectrum-Roentgen-Gamma (SRG) mission. 

% ------------------------------------------------
% ------------------------------------------------

This work presents X-ray ({\it SRG}/eROSITA, {\it Swift}/XRT), optical/UV/mid-infrared (MIR) photometric (ZTF, ASAS-SN, NEOWISE-R, {\it Swift}/UVOT), and optical spectroscopic (NOT/ALFOSC, Las Cumbres Floyds, ANU/WiFeS) observations of AT~2019avd. In Section~\ref{sec:xray_observations}, we report our X-ray observations and analysis of AT~2019avd, whilst the photometric evolution and host galaxy properties are presented in Section~\ref{sec:photo_evolution_host_galaxy}. We then present details of our optical spectroscopic follow-up campaign in Section~\ref{sec:opspec_data_analysis}, before discussing possible origins for AT~2019avd in Section~\ref{sec:discussion}, and conclude in Section~\ref{sec:conclusions}. We adopt a flat $\Lambda$CDM cosmology throughout this paper, with $H_0=67.7\, \mathrm{km}\, \mathrm{s}^{-1}\mathrm{Mpc}^{-1}$, $\Omega _{\mathrm{m}}=0.309$ \citep{Planck2015}; $z=0.029$ thus corresponds to a luminosity distance of 130~Mpc. All magnitudes will be reported in the AB system, unless otherwise stated. 

\section{X-ray observations}\label{sec:xray_observations}
\subsection{eROSITA discovery}
AT~2019avd was discovered in a dedicated search for candidate TDEs in the first eROSITA all-sky survey (eRASS1). Here, the eROSITA source catalogue (version 945 of the source detection pipeline of the eROSITA Science Analysis Software, eSASS, Brunner et al. in prep.) was systematically examined for new soft X-ray sources associated with the nuclei of galaxies that showed no prior indication of being an AGN.

The eROSITA data for AT~2019avd are composed of four consecutive scans with gaps of 4\,hr each and a midtime of 2020-04-28. The total on-source exposure amounts to 140\,s (see Table~\ref{tab:erosita_swift_log}). The source was localised to (RA$_\mathrm{J2000}$, Dec$_\mathrm{J2000})$=(08h23m37s, 04$^{\circ}$23$^\prime$03$^{\prime\prime}$), with a 1$\sigma$ positional uncertainty of 2$^{\prime\prime}$, which is consistent with the nucleus of the galaxy 2MASX J08233674+0423027.

Photons were extracted using the eSASS task SRCTOOL (version 945) choosing a circular aperture of radius 36$^{\prime \prime}$ centred on the above position (84 counts were detected within this region). Background counts were selected from a circular annulus of inner and outer radii 72$^{\prime \prime}$ and 144$^{\prime \prime}$, respectively. Using the best-fit spectral model (see Section~\ref{sec:x-ray_spec_fitting}), we derived a $0.2-2$~keV flux of (1.4$\pm$0.2)$\times 10^{-12}$ erg~cm$^{-2}$s$^{-1}$ (1$\sigma$).

\begin{figure}
    \centering
    \includegraphics[scale=0.8]{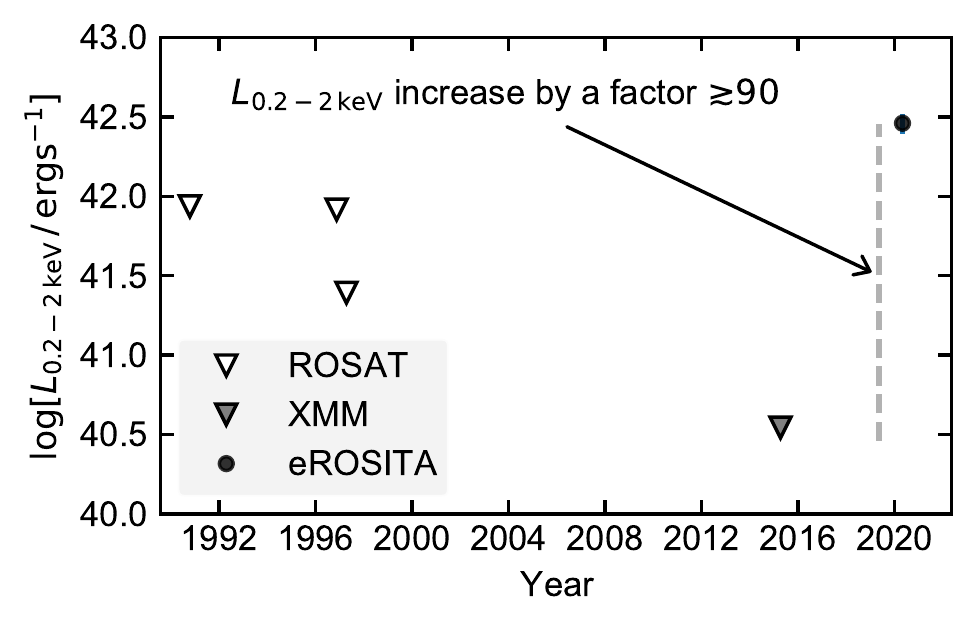}
    \caption{Long-term X-ray light curve in the 0.2--2\,keV energy band of AT~2019avd up until the first eROSITA observation. Triangles denote 3$\sigma$ upper limits for {\it ROSAT}/PSPC and {\it XMM-Newton}/EPIC-pn, whilst the black circle marks the {\it SRG}/eROSITA discovery, where AT~2019avd is at least 90 times brighter than the \textit{XMM-Newton} 3$\sigma$ upper limit. The error bar on the eROSITA marker encloses the 68\% credible region on the observed luminosity.}
    \label{fig:luminosity_history}
\end{figure}

No X-ray source has previously been detected at the location of AT~2019avd. Using both the Upper Limit Server\footnote{\url{http://xmmuls.esac.esa.int/upperlimitserver/}} and webPIMMS\footnote{\url{https://heasarc.gsfc.nasa.gov/cgi-bin/Tools/w3pimms/w3pimms.pl}}, and assuming an absorbed black-body spectral model with $kT=80$\,eV, and Galactic neutral hydrogen column density (see also Section~\ref{sec:x-ray_spec_fitting}), $N_{\mathrm{H}}=2.42\times 10^{20}$ cm$^{-2}$, we infer an $0.2-2$~keV 3$\sigma$ upper limit of 1.7$\times 10^{-14}$ erg~cm$^{-2}$s$^{-1}$ for a serendipitous 7~ks {\it XMM-Newton} pointed observation obtained on 2015-04-08\footnote{\textit{XMM-Newton} OBSID=0741580501}. Earlier constraints can be derived from {\it ROSAT} observations obtained on 1990-10-14, 1996-11-13, and 1997-04-11 with 3$\sigma$ upper limits of 4.2$\times 10^{-13}$, 4.0$\times 10^{-13}$ , and 1.2$\times 10^{-13}$ erg~cm$^{-2}$s$^{-1}$, respectively. 

eROSITA thus first observed AT~2019avd in a state where it had brightened by {at least} a factor of 90 in the $0.2-2$~keV band relative to the deepest archival X-ray observation (luminosity history presented in Fig.~\ref{fig:luminosity_history}).

% ------------------------------------------------
% ------------------------------------------------

\subsection{Swift follow-up}
Triggered by the eROSITA detection, a series of follow-up observations were performed with the \textit{Neil Gehrels Swift Observatory} (P.I.s: A. Malyali \& B. Trakhtenbrot). Observations were obtained roughly every 7 days, until the source was no longer visible due to Sun angle constraints; a further \textit{Swift} observation was then obtained $\sim 3$ months later. A log of the observations can be found in Table~\ref{tab:erosita_swift_log}. The XRT observations were performed in photon counting mode. The data were reduced using the \textsc{xrtpipeline} task included in version 6.25 of the \textsc{heasoft} package. Spectra for each of the five epochs were extracted using the \textsc{xrtproducts} task. Source counts were extracted from a circular aperture of radius 47$^{\prime \prime}$ and background counts extracted from a circular annulus of inner and outer radii 70$^{\prime \prime}$ and 250$^{\prime \prime}$, respectively\footnote{eROSITA and XRT have different PSFs and instrument backgrounds, thus the radii of the extraction regions were chosen based on each instrument and differ here.}. 

% ------------------------------------------------
\begin{table}
        \centering
        \caption{Log of {\it SRG}/eROSITA and {\it Swift}/XRT observations of AT~2019avd until 2020-09-16. For eROSITA, the mid-date of the coverage in eRASS1 is given.}
        \label{tab:erosita_swift_log}
        \begin{tabular}{ccccc} % four columns, alignment for each
                \hline
                Date & MJD & Telescope & ObsID & Exp. [s] \\
                \hline
                2020-04-28 & 58967.7 & {\it SRG}/eROSITA & - & 140  \\
                2020-05-13 & 58982.4 & {\it Swift}/XRT & 00013495001 & 1617  \\
                2020-05-19 & 58988.3 & {\it Swift}/XRT  & 00013495002 & 1966 \\
                2020-05-25 & 58994.0 & {\it Swift}/XRT  & 00013495003 & 1982  \\
                2020-06-03 & 59003.3 & {\it Swift}/XRT  & 00013495004 & 494  \\
                2020-06-10 & 59010.6 & {\it Swift}/XRT  & 00013495005 & 1739 \\
                2020-09-16 & 59108.4 & {\it Swift}/XRT  & 00013495006 & 2967 \\
                \hline
        \end{tabular}
\end{table}
% ------------------------------------------------

Observations with the Ultraviolet and Optical Telescope (UVOT; \citealt{Roming2005}) were obtained simultaneously with the XRT observations. Imaging was performed at three epochs (00013495001, ..004, ..005) using the UVW1 filter with exposures of 1.36, 1.95, and 1.93\,ks, respectively. The remaining three observations utilised all six UVOT filters (UVW2, UVM2, UVW1, U, B, V) with accordingly shorter exposure times.

The UVOT flux was extracted with the \textsc{uvotsource} task using a $9^{\prime \prime}$ radius aperture centred on the optical position of AT~2019avd, whilst a nearby circular region with $15^{\prime \prime}$ radius was used for background subtraction. The photometry was extracted from each unique \textit{Swift} observation ID, and is presented in Table~\ref{tab:UVOTphotometry} (we note that this photometry includes both AGN and host galaxy emission in order to be consistent with the SED fitting in Section~\ref{sec:host_galaxy_properties}). Relative to UV photometry obtained prior to the initial optical outburst (see Section~\ref{sec:host_galaxy_properties} and Fig.~\ref{fig:sed_fit}), AT~2019avd has brightened by $\sim 1$ mag in the UVW1, UVM2, and UVW2 bands, and brightens only by $\sim 0.1-0.2$~mag over \textit{Swift} observations between 2020-05-13 and 2020-09-16. 

\begin{table}
\centering
\caption{\textit{Swift} UV photometry (corrected for Galactic extinction using the UVOT correction factors in Table~5 of \citealt{Kataoka2008}). The model magnitudes (for the host galaxy) were obtained by convolving the best-fit SED model (Section~\ref{sec:host_galaxy_properties}) with the UVOT transmission curves. A hyphen denotes that the given filter was not used on that observation date.}
\label{tab:UVOTphotometry}
\begin{tabular}{rccccccc}
\hline
\hline
Date & UVW1 & UVM2 & UVW2 \\
\hline
Model & 18.88 & 19.16 & 19.26 \\
2020-05-13 & $ 18.01 \pm 0.04$ & - & - \\
2020-05-19 & $ 18.23 \pm 0.15$ & $ 18.28 \pm 0.11$ & $ 18.27 \pm 0.10$ \\
2020-05-25 & $ 17.85 \pm 0.07$ & $ 18.30 \pm 0.07$ & $ 18.31 \pm 0.06$ \\
2020-06-03 & $ 17.89 \pm 0.04$ & - & - \\
2020-06-10 & $ 17.80 \pm 0.04$ & - & - \\
2020-09-16 & $ 17.78 \pm 0.05$ & $ 18.17 \pm 0.06$ & $ 18.23 \pm 0.05$ 
\\
\hline
\end{tabular}
\end{table}

\subsection{X-ray spectral fitting}
\label{sec:x-ray_spec_fitting}
\begin{figure}
    \centering
    \includegraphics[scale=0.8]{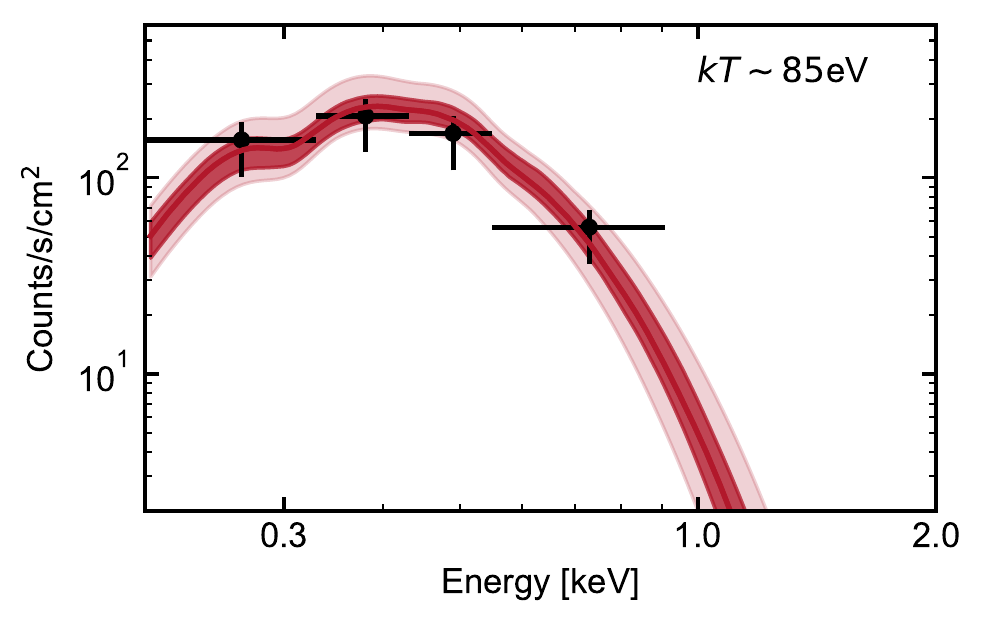}
    \caption{BXA fit to the eROSITA eRASS1 spectrum. Black markers are the binned observed data, whilst the red represents the fitted convolved model for \texttt{tbabs*blackbody} (darker and light red bands enclose the 68\,\% and 95\,\% posterior uncertainty on the model at each energy). Both the black-body and power-law fits to the (low count) eRASS1 spectrum suggest that the source is ultra-soft (see Table~\ref{tab:bxa_results}).}
    \label{fig:convolved_posterior}
\end{figure}

X-ray spectra were analysed using the Bayesian X-ray Analysis software (BXA, \citealt{Buchner2014X-rayCatalogue}), which connects the nested sampling algorithm MultiNest \citep{Feroz2008} with the fitting environment CIAO/Sherpa \citep{10.1117/12.447161} and XSPEC \citep{1996ASPC..101...17A}. The spectra were fitted unbinned using the C-statistic \citep{Cash1976}, and the eROSITA and XRT backgrounds were both modelled using the principal component analysis (PCA) technique described in \citet{Simmonds2018}. For each set of eROSITA and XRT spectra, a joint fit on both the source and background spectra was run. Two different models for the source spectra were used: (i) an absorbed black body (\texttt{tbabs*blackbody}), and (ii) an absorbed power law (\texttt{tbabs*powerlaw}). The equivalent Galactic neutral hydrogen column density, $N_{\mathrm{H}}$, was allowed to vary by 20\% from its tabulated value in the HI4PI survey of $2.42\times10^{20}$\,cm$^{-2}$ \citep{Bekhti2016} during fitting. 
The complete set of priors adopted under each model is listed in Table~\ref{tab:xspec_priors}, whilst an example of the BXA fit to the eROSITA spectrum is shown in Fig.~\ref{fig:convolved_posterior}, and spectral fit results are presented in Table~\ref{tab:bxa_results}.
\begin{table*}
        \centering
        \caption{Summary of priors adopted in the BXA analysis of the eROSITA and XRT spectra. For each fit, a log-uniform prior on $N_{\mathrm{H}}$ between $(0.8N_{\mathrm{H}}, 1.2N_{\mathrm{H}})$ was defined, where $N_{\mathrm{H}}=2.42\times10^{20}$\,cm$^{-2}$ (see Section~\ref{sec:x-ray_spec_fitting}). 
        $\Gamma$ denotes the slope of a power law, $kT$ the black-body temperature, $A$ the normalisation. The prior over $A$ is in units 1.05$\times 10^{-6}\mathrm{erg}\,\mathrm{cm}^{-2}\mathrm{s}^{-1}$.}  
        \label{tab:xspec_priors}
        \begin{tabular}{lccr} 
                \hline
                 Model & Priors \\
                \hline
                \texttt{tbabs*bbody} & $\log [kT / \mathrm{keV}] \sim \mathcal{U}(-2,1)$, $\log [A]\sim \mathcal{U}(-10, 10)$ \\ \texttt{tbabs*powerlaw} & $\Gamma \sim \mathcal{U}(0,8)$, $\log [A]\sim \mathcal{U}(-10, 10)$   \\
                \hline
        \end{tabular}
\end{table*}

\begin{table*}
\centering
\caption{X-ray spectral fit results from applying BXA to the extracted eROSITA and XRT spectra, with uncertainties enclosing 68\% of the posterior for each parameter. $F_{\rm 0.2-2keV}$ is the inferred observed (unabsorbed) flux under each model.}
\label{tab:bxa_results}
\begin{tabular}{r|ccccccc}
\hline
OBSID & \multicolumn{3}{c}{\texttt{tbabs*blackbody}} & \multicolumn{3}{c}{\texttt{tbabs*powerlaw}} \\
\hline
& N$_{\rm H}$ & kT & $F_{\rm 0.2-2keV}$ & N$_{\rm H}$ & $\Gamma$ & $F_{\rm 0.2-2keV}$ \\
& [$\times10^{20}$cm$^{-2}$] & [eV] & [$\times10^{-12}$\,erg cm$^{-2}$ s$^{-1}$] & [$\times10^{20}$cm$^{-2}$] & & [$\times10^{-12}$\,erg cm$^{-2}$ s$^{-1}$] \\
\hline
eRASS1 & $2.3_{\rm -0.3}^{\rm +0.3}$ & $85_{\rm -5}^{\rm +6}$ & $1.4_{\rm -0.2}^{\rm +0.2}$ & $2.5_{\rm -0.3}^{\rm +0.3}$ & $4.2_{\rm -0.3}^{\rm +0.3}$ & $1.6_{\rm -0.2}^{\rm +0.2}$ & \\
00013495001 & $2.4_{\rm -0.3}^{\rm +0.4}$ & $72_{\rm -8}^{\rm +8}$ & $1.4_{\rm -0.2}^{\rm +0.2}$ & $2.4_{\rm -0.3}^{\rm +0.3}$ & $5.3_{\rm -0.4}^{\rm +0.4}$ & $2.5_{\rm -0.5}^{\rm +0.5}$ \\
00013495002 & $2.4_{\rm -0.3}^{\rm +0.3}$ & $83_{\rm -11}^{\rm +12}$ & $1.4_{\rm -0.4}^{\rm +0.4}$ & $2.4_{\rm -0.3}^{\rm +0.3}$ & $5.2_{\rm -0.6}^{\rm +0.7}$ & $2.6_{\rm -0.8}^{\rm +0.8}$ \\
00013495003 & $2.4_{\rm -0.3}^{\rm +0.3}$ & $132_{\rm -10}^{\rm +10}$ & $1.0_{\rm -0.1}^{\rm +0.1}$ & $2.5_{\rm -0.3}^{\rm +0.3}$ & $3.7_{\rm -0.3}^{\rm +0.2}$ & $1.4_{\rm -0.2}^{\rm +0.2}$ \\
00013495004 & $2.4_{\rm -0.3}^{\rm +0.3}$ & $107_{\rm -10}^{\rm +10}$ & $1.0_{\rm -0.2}^{\rm +0.2}$ & $2.4_{\rm -0.3}^{\rm +0.3}$ & $4.2_{\rm -0.3}^{\rm +0.3}$ & $1.6_{\rm -0.3}^{\rm +0.3}$ \\
00013495005 & $2.4_{\rm -0.3}^{\rm +0.3}$ & $91_{\rm -6}^{\rm +6}$ & $1.5_{\rm -0.2}^{\rm +0.2}$ & $2.5_{\rm -0.3}^{\rm +0.3}$ & $4.9_{\rm -0.3}^{\rm +0.3}$ & $2.6_{\rm -0.4}^{\rm +0.4}$ \\
00013495006 & $2.4_{\rm -0.3}^{\rm +0.3}$ & $115_{\rm -3}^{\rm +3}$ & $9.7_{\rm -0.4}^{\rm +0.4}$ & $2.8_{\rm -0.1}^{\rm +0.1}$ & $4.1_{\rm -0.1}^{\rm +0.1}$ & $14.0_{\rm -0.7}^{\rm +0.7}$ \\
\hline
\end{tabular}
\end{table*}

\begin{figure}
    \centering
    \includegraphics[scale=0.8]{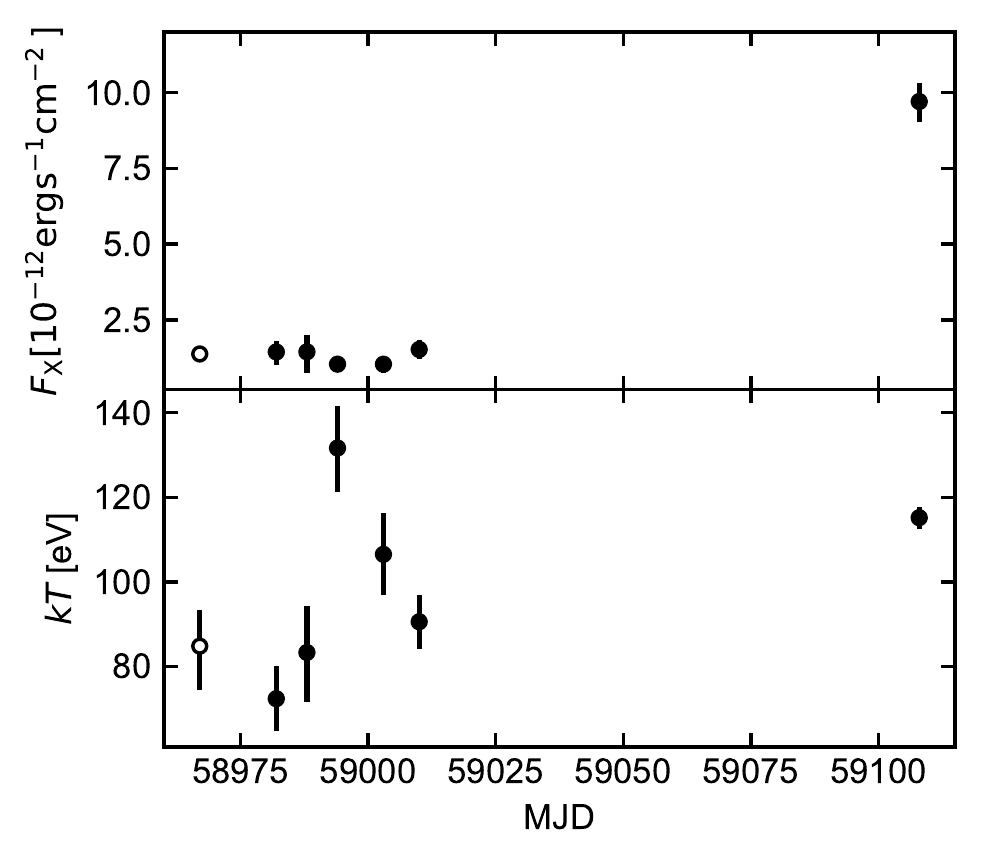}
    \caption{X-ray evolution of AT~2019avd. The empty and filled black markers represent the eROSITA and XRT observations respectively; error bars enclose 95\% of the posterior.}
    \label{fig:xray_evolution}
\end{figure}

Over the course of the six weeks following the initial eROSITA detection, there was no major variability in the $0.2-2$\,keV flux between the eROSITA and XRT observations (Table~\ref{tab:bxa_results} and Fig.~\ref{fig:xray_evolution}). However, the $0.2-2$\,keV flux in the last \textit{Swift} epoch increased by a factor of about six relative to the previous observation.

AT~2019avd remained in an ultra-soft state during the \textit{Swift} monitoring campaign, although there is variability in the inferred black-body temperatures ($kT$ ranges between minimum and maximum values of 72$\pm 8$\,eV and 132$\pm 10$\,eV, respectively). The inferred black-body temperatures are similar to those measured in the X-ray emission of previously observed thermal TDEs ($45\lesssim kT \lesssim 130$\,eV, e.g. \citealt{van2020seventeen}), and are also consistent with the temperatures of the soft excess shown in AGN (e.g. Table~A1 in \citealt{Gliozzi2020}).

\section{Photometric evolution and host galaxy properties}\label{sec:photo_evolution_host_galaxy}
\subsection{Optical evolution}\label{sec:optical_observations}
The region around the position of AT~2019avd has been monitored by ZTF \citep{Bellm2019,Graham2019} in the $r$ and $g$ bands from 2019-01-12 until the time of writing. On 2019-02-09 (over a year before the eROSITA detection), ZTF first detected the transient ZTF19aaiqmgl with an inferred separation from the galaxy centre of 0\farcs04 \footnote{\url{https://lasair.roe.ac.uk/object/ZTF19aaiqmgl/}}, and $r$-band magnitude $17.64 \pm 0.07$ (reference subtracted, Fig.~\ref{fig:finder_chart}). 

\begin{figure*}
    \centering
    \includegraphics[scale=0.8]{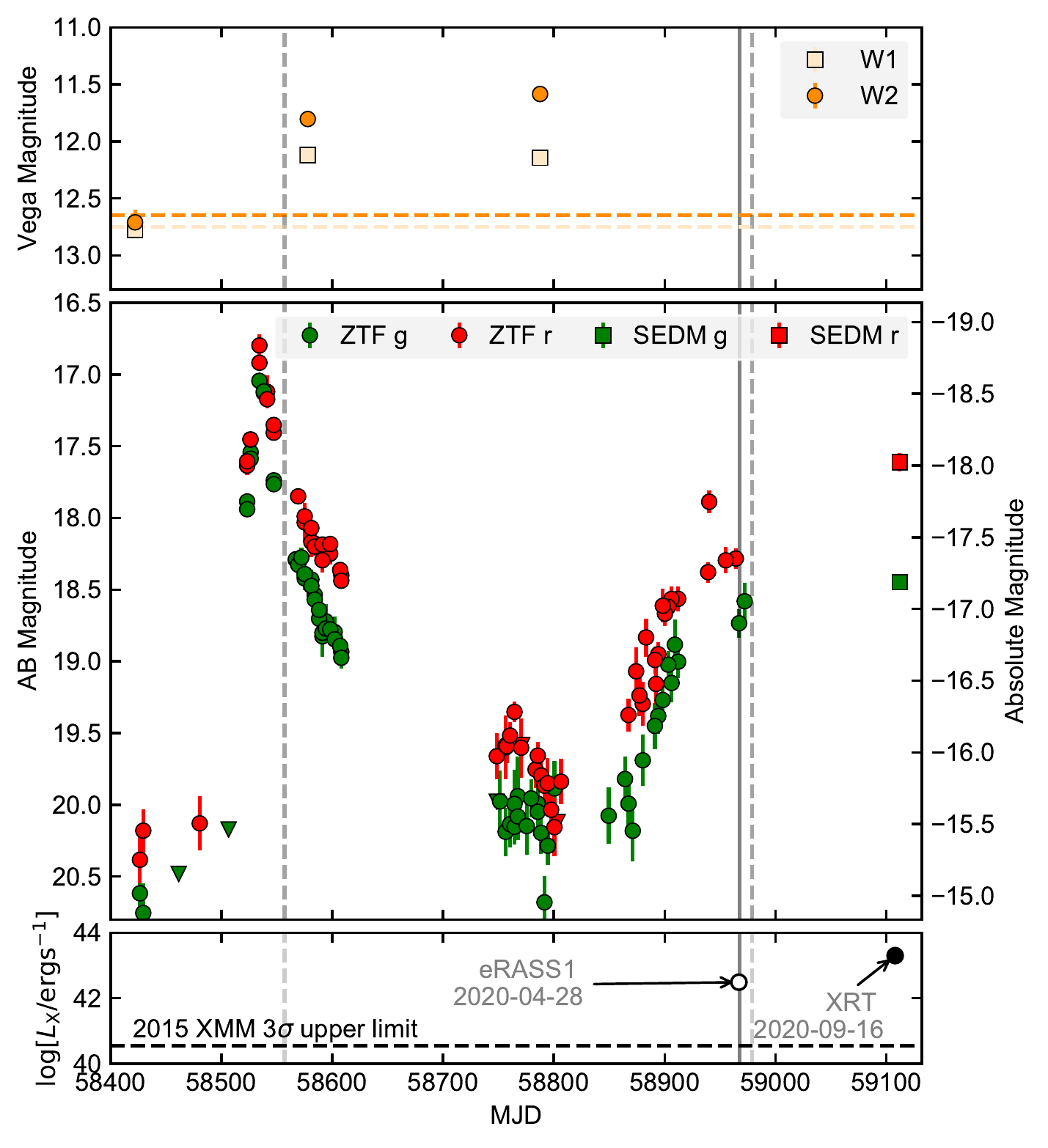}
    \caption{NEOWISE-R (non-host subtracted, top) and ZTF/ SEDM (middle) light curves of  AT~2019avd, with the immediate $0.2-2$~keV X-ray history shown in the bottom panel. The eROSITA eRASS1 detection and the \textit{Swift} observation from 2020-09-16 are the empty and filled black markers, respectively. The solid grey vertical line marks the MJD of the eRASS1 observation, whilst grey dashed lines mark the times of the NOT and the first FLOYDS spectrum (Table~\ref{tab:spectroscopic_log}). No significant variability before the initial 2019 outburst is observed in the host nucleus of AT~2019avd with archival NEOWISE-R and ASAS-SN observations (Fig.~\ref{fig:archival_lightcurve}). The NEOWISE-R observations pre-outburst are observed with mean W1, W2 marked out in the top panel by the cream and orange dashed lines respectively. For plotting clarity, we omit the high-cadence ZTF Partnership observations obtained between MJD 58820 and 58860, and we rebin the $\sim 3$ SEDM observations in each filter into a single data point.}
    \label{fig:ztf_lc_zoomed}
\end{figure*}

For MJD$<$58855 (2020-01-07), we obtained a forced photometry ZTF light curve for AT~2019avd \citep{Masci2018}. For MJD$>$58855, we downloaded the ZTF light curve of AT~2019avd using the Lasair alert broker \citep{Smith_2019}, which processes and reports to the community on transients detected within the large ZTF data streams. Both of these light curves are constructed from PSF-fit photometry measurements run on ZTF difference images. We also obtained additional photometric observations with the Spectral Energy Distribution Machine (SEDM; \citealt{Blagorodnova2018}) on the Palomar 60-inch telescope. The SEDM photometry was host-subtracted using SDSS reference images, as described in \citet{Fremling2016}. 
These two light curves, and the host-subtracted SEDM photometry, were then combined for subsequent analysis, and are shown in Fig.~\ref{fig:ztf_lc_zoomed}.

After the initial detection on 2019-02-09, AT~2019avd continued to brighten until reaching its maximum observed brightness of $r\sim 16.8$\,mag on 2019-02-20. Between 2019-02-24 and 2020-01-01, the  $g$-band magnitude of the host nucleus decayed nearly monotonically from 17.13$\pm$0.09\,mag to 20.08$\pm$0.20\,mag, followed by a re-brightening to 18.58$\pm$0.13\,mag on 2020-05-03. The late time SEDM photometry around 2020-09-19 revealed a further brightening to $r$ and $g$-band magnitudes of $\sim17.6$~mag and $\sim 18.4$~mag respectively. The first eROSITA observation occurred during the rise of the second major peak of the ZTF light curve (Fig.~\ref{fig:ztf_lc_zoomed}).

The location of AT~2019avd has also been monitored in the $V$-band by ASAS-SN \citep{Shappee2014,Kochanek2017TheV1.0} from February 2012 to November 2018, and in the $g$-band from October 2017 to September 2020 (the time of writing). No major optical outbursts were seen in the ASAS-SN light curve prior to the ZTF detection (Fig.~\ref{fig:archival_lightcurve}); given the joint ASAS-SN and ZTF light curves, it is likely that the system `ignited' around MJD~$=58510$ (2019-01-27). 

\subsubsection{Rise and decay timescales in the light curve}
In the following, we fit the light-curve model presented in equation 1 of \citet{VanVelzen2019}, which models the rise with a half-Gaussian function, and an exponential function for the decay, to the first and second peaks of the ZTF light curve, using UltraNest\footnote{\url{https://github.com/JohannesBuchner/UltraNest}} \citep{Buchner2016,Buchner2019} as our sampler. Whilst such a model is not physically motivated, it enables a comparison of the timescales involved in the light curve of AT~2019avd with those of the population of ZTF nuclear transients presented in \citet{VanVelzen2019}.

\begin{figure}
    \centering
    \includegraphics[scale=0.8]{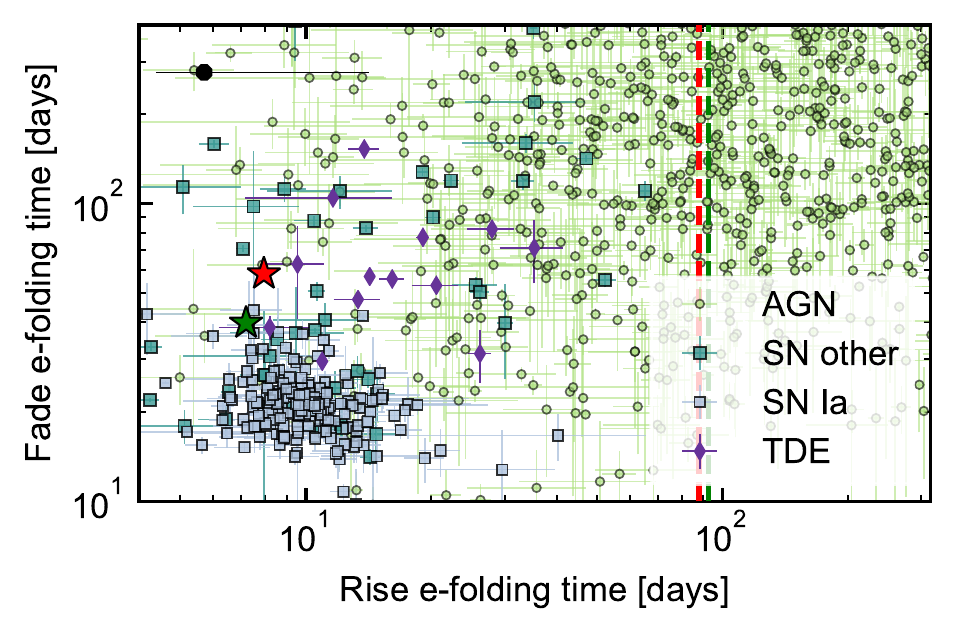}
    \caption{AT~2019avd variability compared with previously classified ZTF nuclear transients (non-AT~2019avd data presented originally in \citealt{van2020seventeen}), with red and green stars computed from the fitted model components for each respective filter. The red and green vertical lines mark the e-folding rise time of the second optical peak in the $r$ and $g$ bands, respectively. We also plot the rise and decay e-fold timescales inferred from the ASAS-SN \textit{V}-band light curve of the nuclear transient AT~2017bgt (\citealt{Trakhtenbrot2019a}; see also Section~\ref{sec:non_tde}) with a black marker. Not only is the double-peaked light curve of AT~2019avd clearly distinct from the other light curves of sources in the AT~2017bgt nuclear transient class, but the first peak of AT~2019avd decays much faster than the AT~2017bgt flare, whilst the second peak rises much slower than the AT~2017bgt flare.}
    \label{fig:optical_lc_fits}
\end{figure}

While fitting the first peak, we first filter out observations outside of the MJD period between 58450 and 58650, and we then run a joint fit of the $g$ and $r$ band observations in flux space. Our model has seven free parameters, defined following the notation of \citet{VanVelzen2019}: $\sigma _r$ and $\sigma _g$, the rise timescale of the light curve in the $r$ and $g$ bands respectively; $\tau _r$ and $\tau _g$, the decay timescale of the light curve in $r$ and $g$ bands; $F_{\mathrm{peak}, r}$ and $F_{\mathrm{peak}, g}$, the peak flux in $r$ and $g$ bands; $t_{\mathrm{peak}}$, the time of the peak of the light curve (to enable a comparison with \citealt{VanVelzen2019}, we assume that the light-curve model peaks at the same time in both of these bands). 
For the second peak, we filter out observations outside of the MJD period 58840 and 59115 (the late-time SEDM datapoints are used in the fitting), and because we do not sample the decay of this peak, we only model the rise here. The model for the second peak has five free parameters, with $\tau _r$ and $\tau _g$ now being omitted. We list our priors in Table~\ref{tab:lightcurve_priors}, and present the fits in Fig.~\ref{fig:optical_lc_fits}.

From the posterior means, we infer $\sigma _{r}=7.9 \pm 0.3$, $\sigma _{g}=7.2 \pm 0.2$, $\tau _r = 58.2 \pm 0.5$ and $\tau _g = 39.8 \pm 0.4$ days for the first optical peak (68\% credible intervals). Whilst the rise timescales in each filter are consistent with each other to within 2$\sigma$, the decay timescales in each filter significantly differ. With $\tau _r > \tau _g$, the first peak shows a potential cooling signature during its decay phase, although we are unable to constrain the temperature evolution during this because of a lack of contemporaneous observations in other wavelength bands. Relative to the population of nuclear transients in \citet{VanVelzen2019}, one sees that these are short rise and decay timescales relative to those of AGN flares, and are thus more similar to those in the  \citet{VanVelzen2019} sample of TDEs and SNe (Fig.~\ref{fig:optical_lc_fits}). As expected from Fig.~\ref{fig:ztf_lc_zoomed}, the inferred rise times for the second peak are longer and more AGN-like, with $\tau _{r} \sim 88$\, days and $\tau _{g}\sim 93$~days.

\subsection{Mid-infrared variability}
The location of AT~2019avd was observed in the $W1$ (3.4\,$\mu$m) and $W2$ (4.6\,$\mu$m) bands by the Wide-Field Infrared Survey Explorer mission (WISE, \citealt{Wright2010THEPERFORMANCE}) in 2010, Near-Earth Object WISE  (NEOWISE; \citealt{Mainzer2011PRELIMINARYSCIENCE}) in late 2010 and 2011, and from December 2013 until now, twice per year as part of the NEOWISE reactivation mission (NEOWISE-R; \citealt{Mainzer2014INITIALMISSION}). 
The NEOWISE-R light curve was obtained from the NASA/IPAC Infrared Science Archive\footnote{\url{https://irsa.ipac.caltech.edu/frontpage/}} by compiling all source detections within 5$^{\prime \prime}$ of the ZTF transient position. 
Individual flux measurements were  rebinned to one data point per NEOWISE-R all-sky scan (using a weighted mean) and converted into magnitudes. The resulting light curve is shown in Fig~\ref{fig:ztf_lc_zoomed}.

The MIR light curve was observed to be flat prior to the initial ZTF outburst, but showed significant brightening in the first NEOWISE-R epoch obtained thereafter. Observations obtained $~\sim$ 6 months later found the source to
still  be in the bright state despite having faded by $\sim3$\,mag in the optical.  
The MIR brightening was also accompanied by a significant reddening, evolving from $W1-W2 \sim 0.08$~mag in AllWISE, to a more AGN-like $W1-W2 \sim 0.6$~mag during flaring. The $W1-W2$ colour before the outburst is much lower than the suggested cuts ($W1-W2\gtrsim0.7$~mag) for identifying AGNs in previous MIR classification schemes \citep{Stern2012,Assef2013,Assef2018}, further  supporting the hypothesis that there was no strong recent AGN activity in AT~2019avd at that time (although the use of WISE colours for selecting AGNs is less effective at lower AGN luminosities; see discussion in \citealt{Padovani2017}). 

\subsection{Host-galaxy properties}\label{sec:host_galaxy_properties}
\begin{figure*}
    \centering
    \includegraphics[scale=0.8]{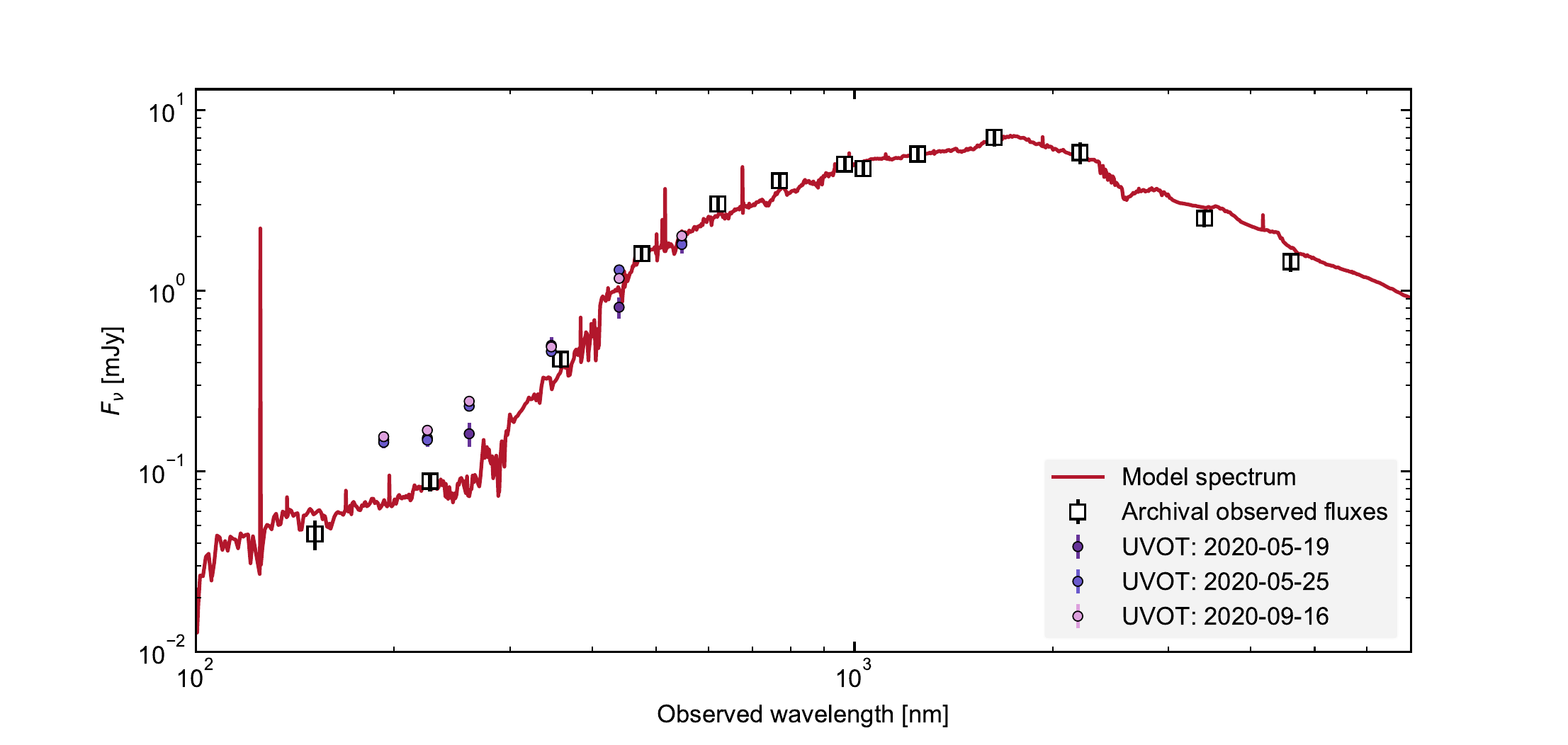}
    \caption{Spectral energy distribution of the host galaxy of AT~2019avd compiled from archival GALEX, SDSS, UKIDSS, and ALLWISE photometry, with the best-fit model shown as a red solid line. The three epochs of \textit{Swift} UVOT photometry where all filters were used are also plotted. AT~2019avd shows a $\sim 1$~mag rise in the UVW1, UVM2, and UVW1 bands relative to the best fit model to the archival photometry.}
    \label{fig:sed_fit}
\end{figure*}

The spectral energy distribution (SED) of the host galaxy of AT~2019avd was compiled from archival\footnote{`Archival' is defined here by photometry taken prior to the initial ZTF optical outburst.} UV to MIR photometry from {\it GALEX} (FUV, NUV), SDSS DR12 ($g$, $r$, $i$, $z$), UKIDSS ($y$, $J$, $H$, $K$), and AllWISE (W1, W2). The SED was modelled using CIGALE \citep{Burgarella2005,Boquien2019}, which allows the estimation of the physical parameters of a galaxy by fitting composite stellar populations combined with recipes describing the star formation history and attenuation. The best-fitting model (see Fig.~{\ref{fig:sed_fit}}) is that of a galaxy with a stellar mass of $(1.6\pm0.8)\times 10^{10} M_{\odot}$, a star formation rate (SFR) of $0.17 \pm 0.05$~$M_{\odot}$yr$^{-1}$, and little attenuation, $\mathrm{E(B-V)}=0.03\pm0.02$\,mag, which experienced a burst of star formation $3.7\pm0.2$\,Gyr ago. The inferred stellar mass and SFR place the host galaxy of AT~2019avd in the `green valley’ between the star-forming main sequence and quenched elliptical galaxies (adopting the green valley definition presented in \citealt{Law-Smith2017a}).

The SED fit suggests that the host galaxy did not show strong signs of nuclear activity prior to the detection of AT~2019avd. This is further supported by the absence of a radio counterpart in the FIRST catalogue \citep{Becker1995} within 30$^{\prime \prime}$ of AT~2019avd, with a catalogue upper detection limit at this position of 0.96~mJy/beam\footnote{\url{http://sundog.stsci.edu/cgi-bin/searchfirst}.}. 

\section{Optical spectral analysis}\label{sec:opspec_data_analysis}
\subsection{Spectroscopic observations}\label{sec:opt_spectroscopy}
On 2019-03-15, $\sim$33 days after the first observed peak in the ZTF light curve, an optical spectrum of AT~2019avd was obtained by \citet{2020ATel13717....1G} with the Alhambra Faint Object Spectrograph and Camera (ALFOSC)\footnote{\href{http://www.not.iac.es/instruments/alfosc}{\url{http://www.not.iac.es/instruments/alfosc}}} on the 2.56\,m Nordic Optical Telescope (NOT). The spectrum was obtained with a 1\farcs0 wide slit, grism \#4 (covering the wavelength region from 3650-9200~{\AA}), and the slit was positioned along the parallactic angle at the beginning of the 1800s exposure. Reductions were performed in a standard way using mainly \texttt{iraf} based software, including bias corrections, flat fielding, wavelength calibration using HeNe arc lamps imaged immediately after the target and flux calibrations using observations of a spectrophotometric standard star.

No further spectra were taken until after eROSITA had detected the large-amplitude soft-X-ray flare from AT~2019avd in late April 2020, which triggered a further five epochs of spectroscopy (dates listed in Table~\ref{tab:spectroscopic_log}) using the 
FLOYDS spectrographs \citep{Brown2013LasNetwork} mounted on the  Las Cumbres Observatory 2m telescopes at Haleakala, Hawaii, and Siding Spring, Australia. Each spectrum was taken with a 3.6ks exposure, using the `red/blu' grism  and a slit width of 2$^{\prime \prime }$. 
The spectra were reduced using PyRAF tasks as described in \citet{Valenti2014}. FLOYDS covers the entire 3500-10000 $\textrm{\AA}$ range in a single exposure by capturing two spectral orders (one red and one blue) simultaneously, yielding $R\sim400$. The different orders are usually merged into a single spectrum using the region between 4900 and 5700 $\textrm{\AA}$, which is present in both the red and blue orders. However, in this case,  in order to avoid erroneous wavelength shifts at the blue edge of the red order (where there are fewer arclines), all FLOYDS spectra were merged using a reduced stitching region of 5400 to 5500 $\textrm{\AA}$\footnote{The most extreme arcline used to calibrate each order is at $\sim$5460 $\textrm{\AA}$.}. This stitching was done manually in Python, by replacing fluxes in that wavelength range with an average of the linear interpolations of the two orders.

In addition, a higher resolution spectrum (R$\sim$3000) was obtained on 2020-05-29 with the Wide Field Spectrograph (WiFeS; \citealp{Dopita2007TheWiFeS,Dopita2010TheReduction}) mounted on the 2.3m ANU telescope at Siding Spring Observatory. We employed the R3000 and B3000 gratings, and obtained
an arc lamp exposure after each target exposure. The total spectral range from
the two gratings is 3500 to 7000 \AA. The data were reduced using the PyWiFeS reduction pipeline \citep{Childress2014}, which produces three-dimensional data (data cubes). These spectra are bias subtracted, flat-fielded, wavelength and flux calibrated, and corrected for telluric absorption. We then extracted the spectra from the slitlets that captured AT~2019avd using the IRAF \citep{Tody1986} task \textsc{apall} which allowed for background subtraction.

\begin{figure*}
    \centering
    \includegraphics[scale=0.8]{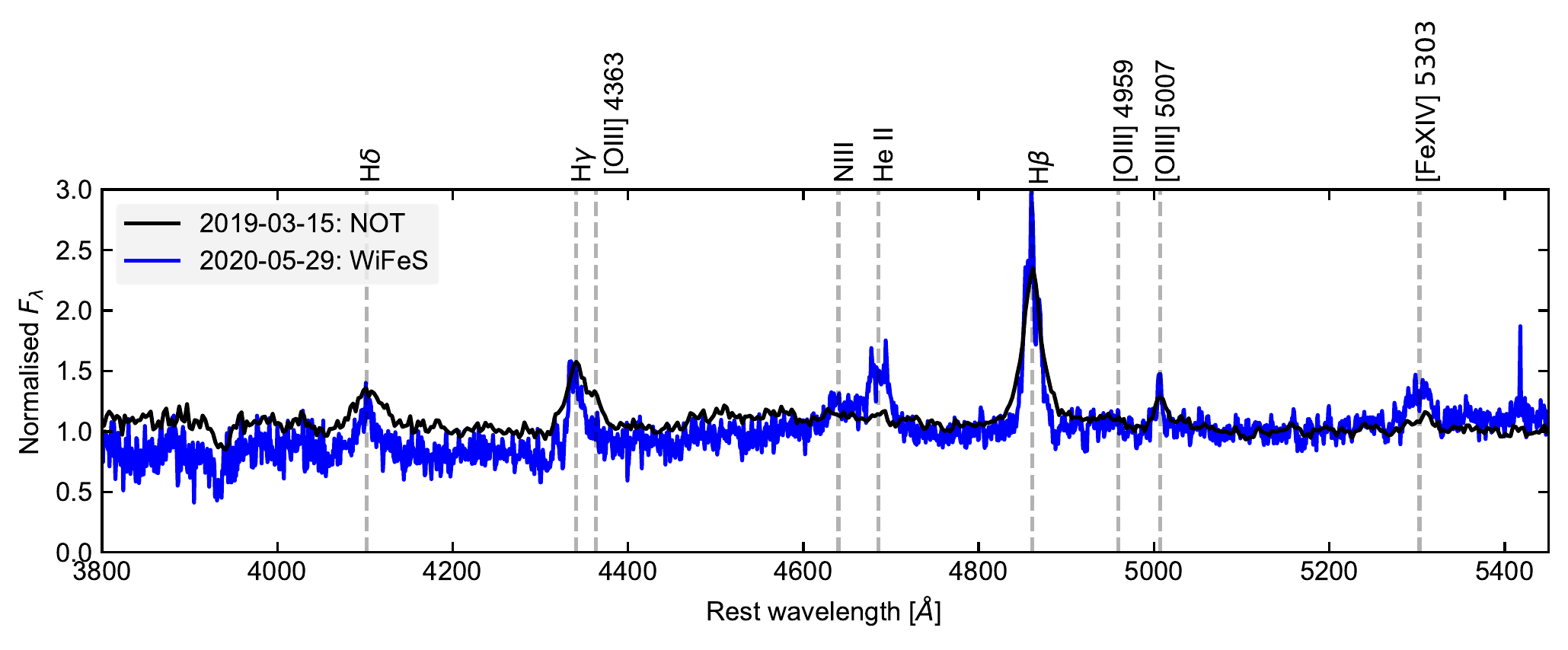}
    \includegraphics[scale=0.8]{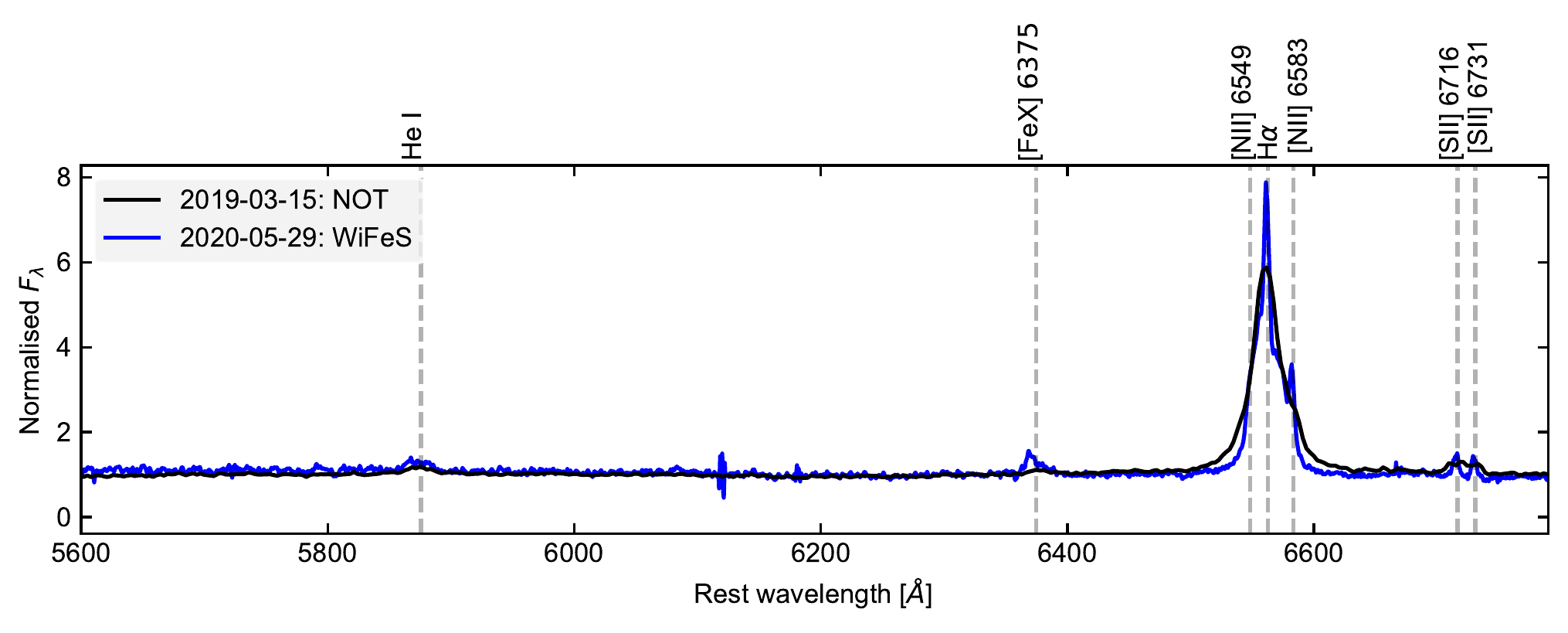}
    \caption{Comparison of NOT and WiFeS spectra (black and blue respectively). The top panel shows the wavelength range 3800-5450 $\textrm{\AA}$, while the bottom panel shows the 5600-6800 $\textrm{\AA}$ range. The most notable changes are (a) the emergence of the broad emission feature around rest-frame wavelength 4686~{\AA} and (b) an increase in intensity of the high-ionisation coronal Fe lines ($\sim$5300 and 6370~{\AA}). The WiFeS spectrum is of much higher resolution relative to the NOT spectrum, and therefore is able to better resolve narrow emission lines, such as the [\ion{S}{II}] doublet at 6716 and 6731~{\AA}. Neither are shown corrected for Galactic extinction. The NOT spectrum was normalised by its continuum flux in the 5100-5200~{\AA} range (rest frame), whilst the blue and red arms of the WiFeS spectra were normalised in the 5100-5200~{\AA} and 6400-6450~{\AA} ranges respectively (rest frame).} 
    \label{fig:not_wifes_comparison}
\end{figure*}

\begin{figure}
    \centering
    \includegraphics[scale=0.8]{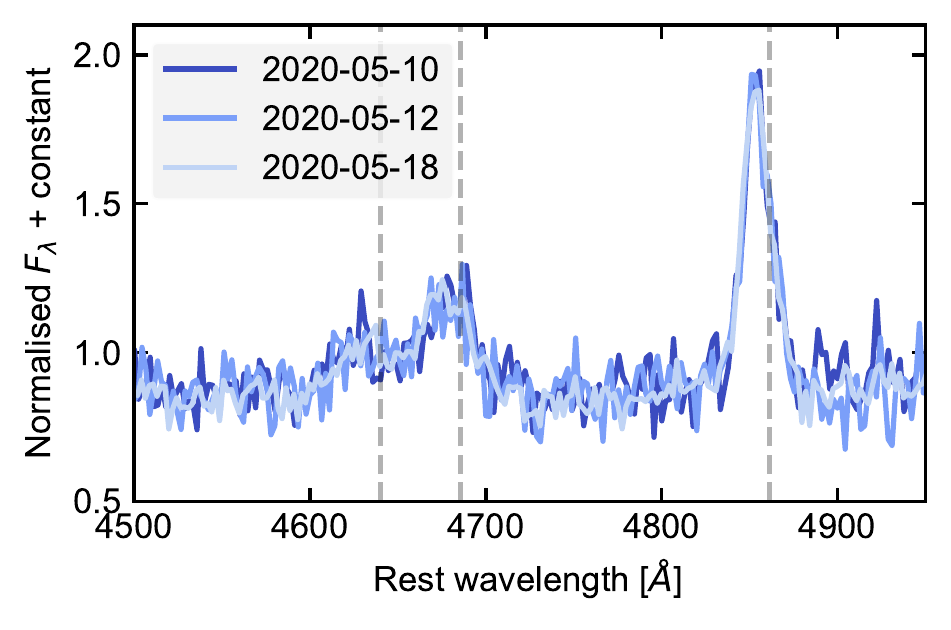}
    \includegraphics[scale=0.8]{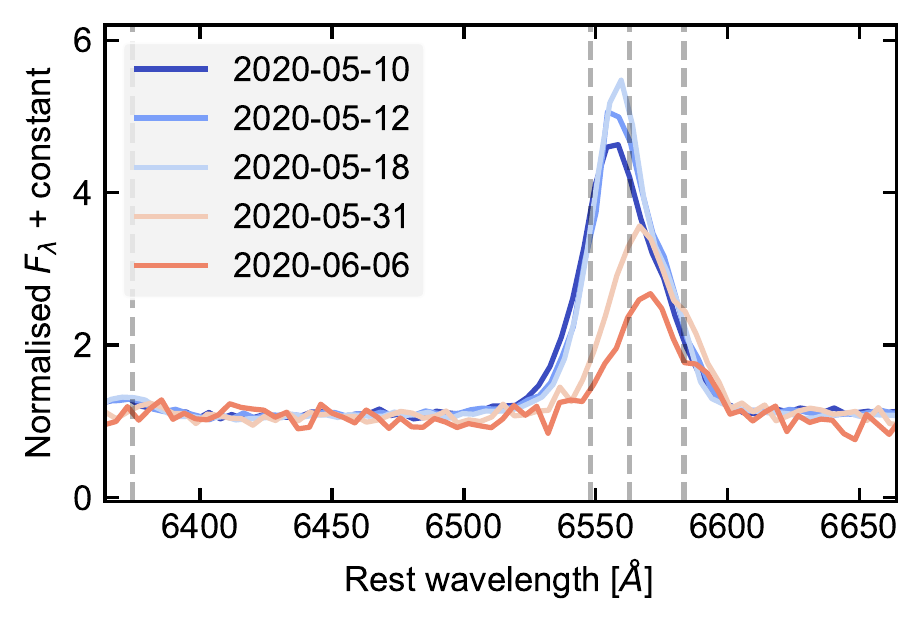}
    \caption{Evolution of the Bowen+H$\beta$ (top) and H$\alpha$ (bottom) Balmer emission lines observed through the five epochs of FLOYDS spectroscopy. Grey dashed lines match those in Fig.~\ref{fig:not_wifes_comparison}. Epochs 2020-05-31 and 2020-06-06 were of low S/N in the blue wavelength range, and thus are omitted from the plot here. The minor evolution of the H$\alpha$ peak position over the FLOYDS spectra was deemed to be most likely due to aperture-related effects during observations.}
    \label{fig:floyds_balmer_evolution}
\end{figure}

A comparison of the NOT and WiFeS spectra is presented in Fig.~\ref{fig:not_wifes_comparison}, and the spectral evolution in the FLOYDS spectra is shown in Fig.~\ref{fig:floyds_balmer_evolution}. A log of the spectroscopic observations of AT~2019avd is presented in Table~\ref{tab:spectroscopic_log}. We note that we have not found any archival optical spectra of the host galaxy that were obtained prior to the initial 2019 outburst discovered by ZTF. 

\begin{table}
        \centering
        \caption{Spectroscopic observations of AT~2019avd.}
        \label{tab:spectroscopic_log}
        \begin{tabular}{cccccc} % four columns, alignment for each
                \hline
                UT Date & Tel. & Instrument & Exp. [ks] & Airmass \\
                \hline
                2019-03-15 & NOT & ALFOSC & 1.8 & 1.5 \\
                2020-05-10 & FTS & FLOYDS-S & 3.6 & 1.4 \\
                2020-05-12 & FTS & FLOYDS-S & 3.6 & 1.6 \\
                2020-05-18 & FTN & FLOYDS-N & 3.6 & 1.6 \\
                2020-05-29 & ANU & WiFeS & 1.8 & 1.5 \\
                2020-05-31 & FTS & FLOYDS-S & 3.6 & 1.7 \\
                2020-06-06 & FTS & FLOYDS-S & 3.6 & 1.9  \\
                \hline
        \end{tabular}
\end{table}

\subsection{Summary of the main observed features of the optical spectra}
The NOT spectrum from 2019-03-15 appears similar to broad line AGN spectra, showing a relatively flat continuum (in terms of $F_\lambda$) and broad Balmer emission lines (H$\alpha$, H$\beta$, H$\gamma$, H$\delta$; Fig.~\ref{fig:not_wifes_comparison}). However, the strong \ion{Fe}{II} complex that is frequently seen in some AGNs is not present. The H$\alpha$ profile is asymmetric due to the blending of unresolved H$\alpha$ and narrow [\ion{N}{II}]~6549,~6583~{\AA} lines, whilst the asymmetry of the H$\gamma$ line is likely due to blending of H$\gamma$ and [\ion{O}{III}] 4363~{\AA} emission. The other notable features are the [\ion{S}{II}] doublet at 6717 and 6731~{\AA} (again blended, but later resolved in the WiFeS spectrum), and the weak He I emission at 5876~{\AA}.  As no archival spectrum of the host galaxy is available, we are unable to judge whether or not the main observed emission features appeared at the onset of the extreme optical variability. The WiFeS spectrum from 2020-05-29 (Fig.~\ref{fig:not_wifes_comparison}) shows the same emission features as the NOT spectrum, with the addition of a broad emission feature around 4680~{\AA} and an apparent increase in intensity of a set of high-ionisation coronal lines ([\ion{Fe}{XIV}] 5303~{\AA} and [\ion{Fe}{X}] 6375~{\AA}, with ionisation potentials of 392 and 262eV respectively). We assume that the [\ion{Fe}{X}] is not blended with the [\ion{O}{I}]~6364~{\AA} emission feature, because the latter is expected to be a third of the intensity of the [\ion{O}{I}]~6300~{\AA} emission (e.g. \citealt{PELAT1987}), which is not detected.

The FLOYDS spectra (Fig.~\ref{fig:floyds_balmer_evolution}) show no major evolution in the Balmer emission line profiles, and show the broad emission feature around 4680~{\AA} from 2020-05-10 (for epochs with sufficiently high S/N ratios in the blue wavelength range), which was reported to the TNS (and first identified) in \citet{2020TNSCR1391....1T}. 

\subsection{Optical spectrum modelling}
\begin{figure*}
    \centering
    \includegraphics[scale=0.8]{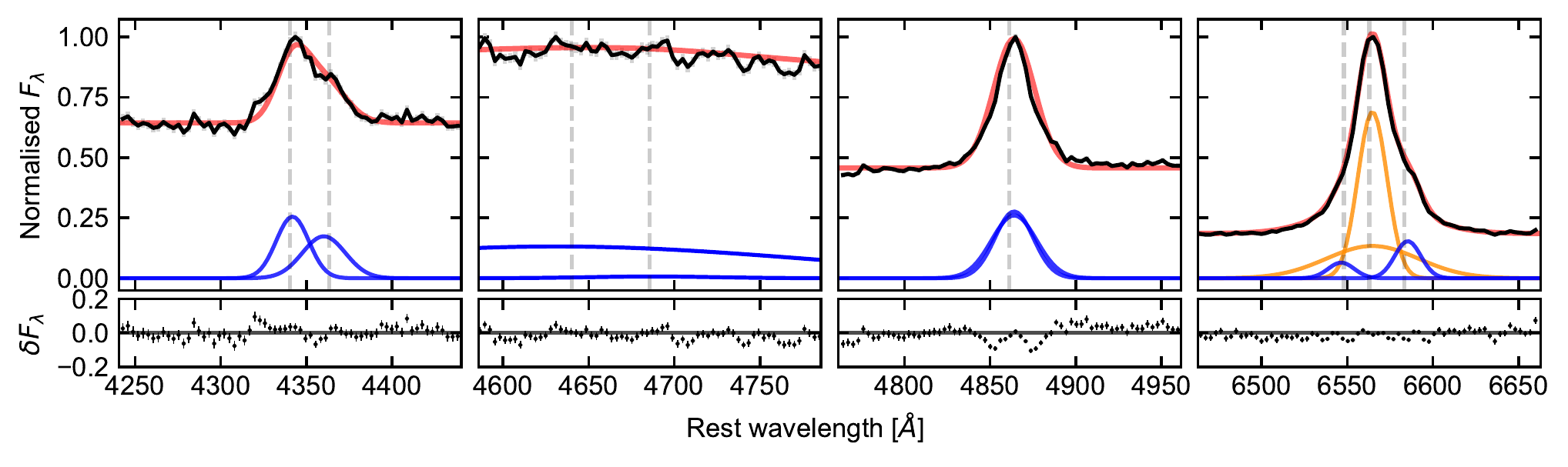}
    \includegraphics[scale=0.8]{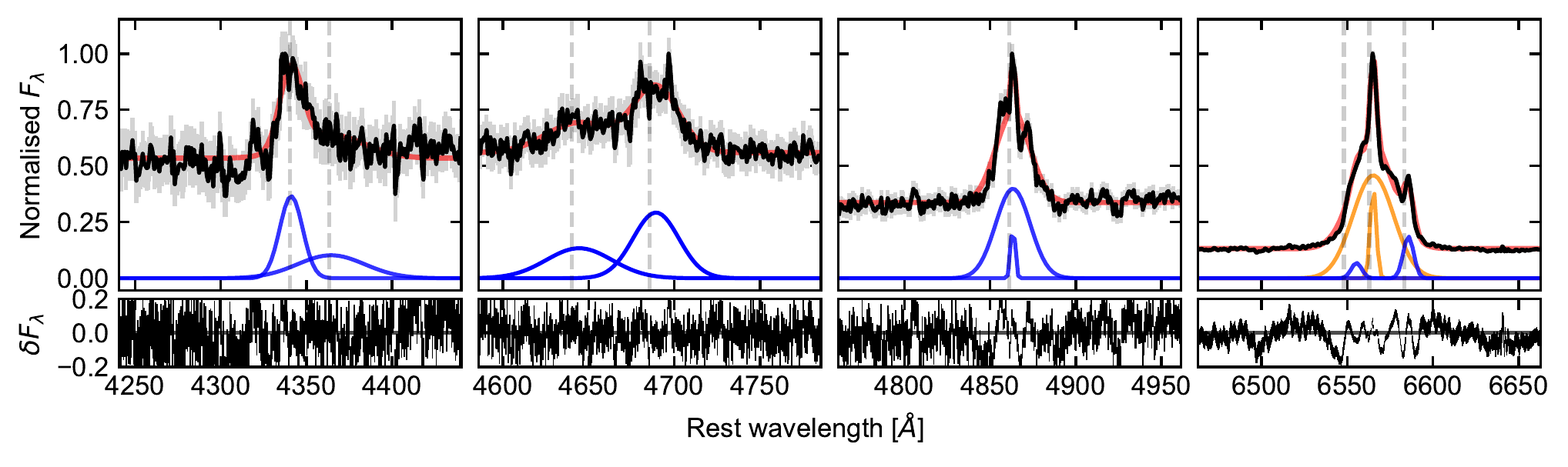}
    \caption{Zoomed-in plots of the main emission lines observed in both the NOT and WiFeS spectra (top and bottom panels respectively). The black line is the observed flux density, and the grey error bars are the associated uncertainties. We plot our fitted spectral model to the data for each region in red (including background component), whilst the blue and orange lines along the bottom represent the  contribution of each source component to the fit (further described in Table~\ref{tab:op_spec_high_res_fit_components}). The lower plots in each panel show the residuals in the spectral fitting, where $\delta F_{\lambda}$ is the difference between the observed $F_{\lambda}$ and the model $F_{\lambda}$, normalised by the model $F_{\lambda}$. We note that the double peaked appearance of the \ion{He}{II} emission line in the WiFeS spectrum is most likely non-physical and due to the noisy optical spectrum, as no other broad lines show such similar line profiles.}
    \label{fig:high_res_spectra_best_fit_models}
\end{figure*}

\begin{figure}
    \centering
    \includegraphics[scale=0.8]{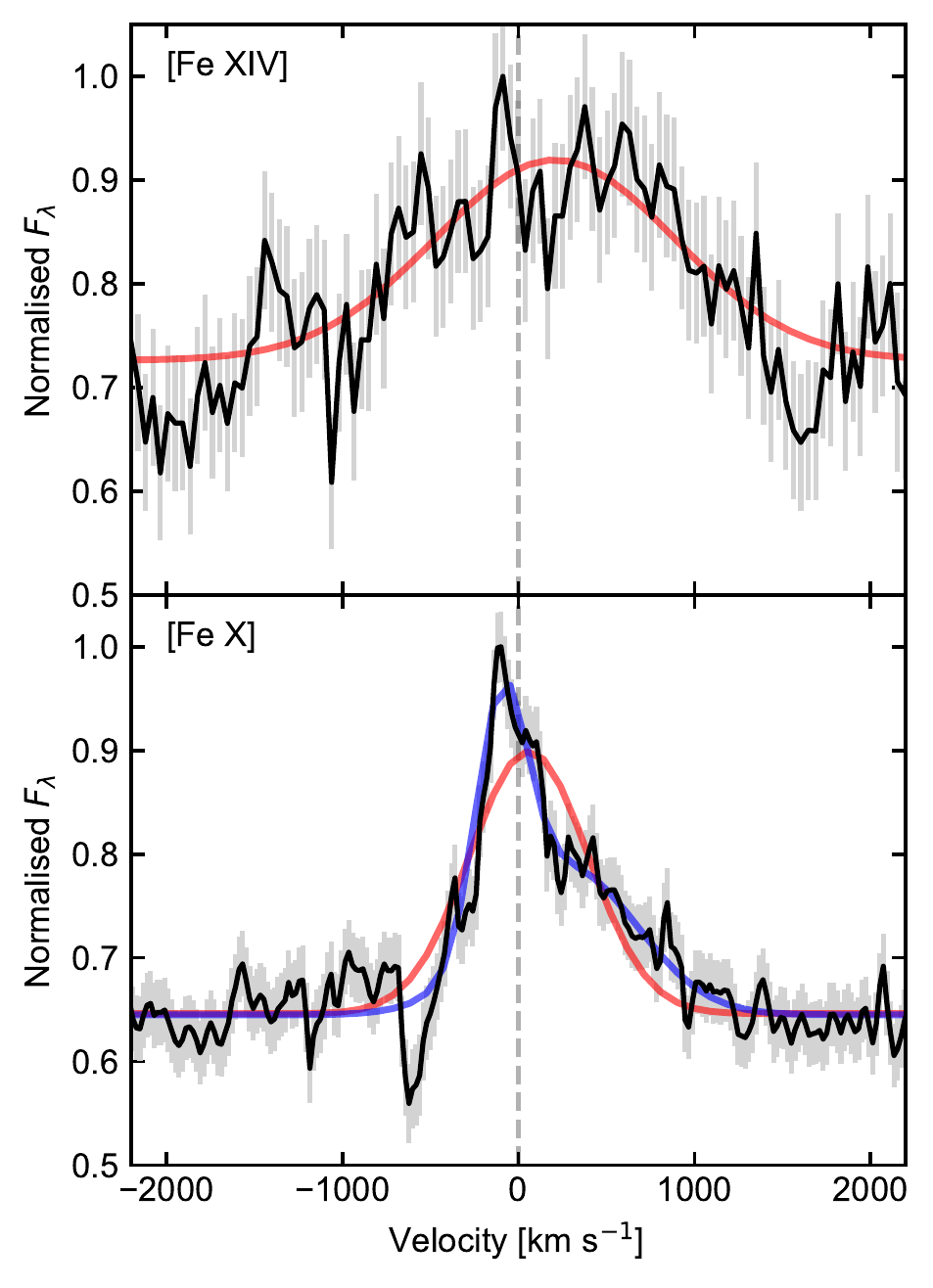}
    \caption{Best-fit single Gaussians (red) to the transient [\ion{Fe}{XIV}] 5303~{\AA} (top) and [\ion{Fe}{X}] 6375~{\AA} (bottom) coronal lines observed in the WiFeS spectrum. The lower ionisation line of the pair, [\ion{Fe}{X}]~6375~{\AA}, is more asymmetric, its broad base appears slightly blueshifted, and can also be fitted by a pair of Gaussians of FWHMs $330 \pm 40$~km~s$^{-1}$ and $900 \pm 100$~km~s$^{-1}$ (blue line), with $F([\ion{Fe}{X}]~6375)/F([\ion{O}{III}]~5007) \sim 2.6$.}
    \label{fig:coronal_line_fits}
\end{figure}

For the two higher resolution spectra (NOT and WiFeS), the region around the main observed emission lines is fitted separately (H$\gamma, 4240{\AA}<\lambda <4440${\AA}; \ion{He}{II}, $4500{\AA}<\lambda <4800{\AA}$; H$\beta, 4700{\AA}<\lambda <5000{\AA}$; H$\alpha, 6364${\AA} $< \lambda <6764{\AA}$; [\ion{S}{II}] doublet, $6650{\AA}<\lambda <6800{\AA}$; and $\pm 100${\AA} of the line centre for [\ion{O}{III}] 5007~{\AA}, [\ion{Fe}{X}] 6375 {\AA}). Each emission line complex is modelled with multiple Gaussians (an overview of these is presented in Table~\ref{tab:op_spec_high_res_fit_components}), and each complex is fitted independently of the others. For all spectral fits, we assume a flat continuum component during the fitting process, and run our model fitting using the region slice sampler option within UltraNest. Spectral fits for the NOT and WiFeS spectra are shown in Figs.~\ref{fig:high_res_spectra_best_fit_models} and \ref{fig:coronal_line_fits}, whilst the spectral fit results are listed in Tables~\ref{tab:optSpecLineRatios}, \ref{tab:optSpecLineRatios_wifes}, and \ref{tab:optSpecFWHM_wifes}.

\begin{table*}
\centering
\caption{Emission line ratios relative to [\ion{O}{III}]~5007{\AA}, where the inferred [\ion{O}{III}]~5007{\AA} flux in each spectrum is $1.34\pm0.09 \, \times 10^{-15} \, \mathrm{erg}\,\mathrm{cm}^{-2}\mathrm{s}^{-1}$ and $4.8\pm0.7 \times 10^{-16} \, \mathrm{erg}\,\mathrm{cm}^{-2}\mathrm{s}^{-1}$. The two spectra were obtained with different slit widths and orientations, and have not been calibrated with independent photometric measurements, hence the line ratios relative to [\ion{O}{III}]~5007{\AA}  reported here. A dashed entry indicates that a given emission line was not clearly detected in the optical spectral fitting.}
\label{tab:optSpecLineRatios}
\begin{tabular}{rcccccc}
\hline
Date & \ion{N}{III} $4640$& \ion{He}{II} $4686$ & H$\rm{\beta}$ & H$\rm{\alpha}$& [\ion{N}{II}] $(6549+6583$) & [\ion{S}{II}] $(6716+6731$)\\
\hline
2019-03-15 & - & - & $10_{\rm -3}^{\rm +3}$ & $38_{\rm -3}^{\rm +3}$ & $7_{\rm -1}^{\rm +1}$ & $1.7_{\rm -0.1}^{\rm +0.1}$  \\
2020-05-29 & $4_{\rm -3}^{\rm +5}$ & $6_{\rm -5}^{\rm +7}$ & $11_{\rm -2}^{\rm +2}$ & $41_{\rm -6}^{\rm +6}$ & $5_{\rm -1}^{\rm +1}$ & $1.7_{\rm -0.2}^{\rm +0.2}$  \\
\hline
\end{tabular}
\end{table*}

\begin{table}
\centering
\caption{Emission line ratios from the WiFeS spectrum, where the narrow components were resolved. The superscript `b' and `n' denote the broad and narrow components, respectively.}
\label{tab:optSpecLineRatios_wifes}
\begin{tabular}{rc}
\hline
Line 1, Line 2 & F(Line 1)/ F(Line 2) \\
\hline
H$\alpha ^{\mathrm{n}}$, H$\beta ^{\mathrm{n}}$ & $ 5.8 \pm 0.8 $
\\
H$\alpha ^{\mathrm{b}}$, H$\beta ^{\mathrm{b}}$ &  $ 3.4 \pm 0.1 $
\\
\ion{He}{II} 4686, H$\beta ^{\mathrm{b}}$ &  $ 0.6 \pm 0.1 $ 
\\
\ion{N}{III} 4640, H$\beta ^{\mathrm{b}}$ & $ 0.4 \pm 0.1 $ 
\\
{[}\ion{Fe}{X}{]}, {[}\ion{O}{III}{]} 5007 & $ 2.4 \pm 0.3 $ 
\\
{[}\ion{Fe}{XIV}{]}, {[}\ion{O}{III}{]} 5007 & $ 3.0 \pm 0.5 $ 
\\
\hline
\end{tabular}
\end{table}

\begin{table}
\centering
\caption{Line widths inferred from the WiFeS spectrum.}
\label{tab:optSpecFWHM_wifes}
\begin{tabular}{rc}
\hline
Line & FWHM [km s$^{-1}$] \\
\hline
\ion{N}{III} 4640 & $2813\pm{648}$ \\
\ion{He}{II} 4686 & $1959\pm{172}$ \\
H$\beta ^{\mathrm{n}}$ & $173\pm{20}$ \\
H$\beta ^{\mathrm{b}}$ & $1422\pm{11}$ \\
{[}\ion{O}{III}{]} 5007 & $384\pm{80}$ \\
{[}\ion{Fe}{XIV}{]} 5303 & $1558\pm{144}$ \\
{[}\ion{Fe}{X}{]} 6375 & $768\pm{35}$ \\
H$\alpha ^{\mathrm{n}}$ & $182\pm{3}$ \\
H$\alpha ^{\mathrm{b}}$ & $1252\pm{9}$ \\
{[}\ion{N}{II}{]} 6549 & $319\pm{12}$ \\
\hline
\end{tabular}
\end{table}

\subsection{Emission line diagnostics}
\subsubsection{Balmer emission}
From the best-fitting spectral models, we infer a broad Balmer decrement, $F(\mathrm{H} \alpha ^\mathrm{b}) / F(\mathrm{H} \beta ^\mathrm{b})$, of 3.4 in the WiFeS spectrum (we use superscripts `b' and `n' to refer to the broad and narrow components of a given emission line when such are clearly detected). Such a decrement is consistent with what is observed in AGNs (e.g. \citealt{Dong2005,Dong2007a,Baron2016}), and is slightly higher than the predicted value of around 2.74-2.86\footnote{The predicted value is dependent on the assumed gas density and temperature.} for case B recombination \citep{Baker1938} and thus a photoionisation origin. Whilst it was originally thought that the observed distribution in the Balmer decrements above 2.86 may have been due to a mix of collisional excitation and dust reddening in the centre of the host galaxy, several papers have suggested that the fundamental driver for this variance is the reddening (e.g. \citealt{Dong2007a,Baron2016,Gaskell2017}). \citet{Dong2007a} find that after accounting for reddening, the intrinsic distribution of Balmer decrements in AGNs is well described by a log Gaussian of mean 3.06, with a 0.03 dex standard deviation, whilst a recent work by \citet{Gaskell2017} find the intrinsic distribution is 2.72$\pm$0.04, and thus consistent with case B recombination. 

Using these results, and by working on the assumption that the intrinsic Balmer decrement is set by Case B recombination to 2.86, we infer an $E(B-V) \sim 0.17$ and $0.65$ mag from the broad and narrow Balmer emission lines respectively (using the \citealt{Calzetti2000} extinction law)\footnote{Alternatively, the inferred $E(B-V)$ values are 0.10 and 0.59 if we assume that the intrinsic Balmer decrement is 3.06 as in \citet{Dong2007a}.}. We note that the E(B-V) inferred from the Balmer decrement is larger than that inferred from SED fitting, which was performed on photometry that included light emitted from a larger region in the host galaxy than that probed by the Balmer decrement analysis.

\subsubsection{Bowen feature around 4680~{\AA}}\label{sec:bowen_diagnostics}
Both the FLOYDS and the WiFeS spectra show the emergence of a broad emission feature around 4680~{\AA}, which is likely a blend of \ion{He}{II} 4686~{\AA} and \ion{N}{III} 4640~{\AA}. Although this feature overlaps with the 4400-4700~{\AA} region, which can often show prominent \ion{Fe}{II} emission in AGNs, we disfavour an \ion{Fe}{II} origin here on the basis of no strong \ion{Fe}{II} bump being observed from the strongest \ion{Fe}{II} transitions in the 4500-4600~{\AA} or $\sim$5150-5350 {\AA} ranges (e.g. \citealt{Kovacevic2010}). When comparing the WiFeS AT~2019avd spectrum to the composite SDSS quasar spectrum presented in Fig.~2 of \citet{Trakhtenbrot2019a}, which was constructed from about $1000$ SDSS quasars with broad Balmer lines of FWHM~$\sim2000 \, \mathrm{km}\, \mathrm{s}^{-1}$, the \ion{He}{II} emission in AT~2019avd is much stronger relative to the Balmer emission in the AGN composite.

The \ion{N}{III}~4640~{\AA} emission suggests the presence of Bowen fluorescence \citep{Bowen1928}. \ion{He}{II} Ly$\alpha$ photons at 303.783~{\AA} are produced after recombination of He$^{++}$ \footnote{The \ion{He}{II} ionisation potential is 54.4~eV.}, and can then either escape, ionise neutral H or He, or, because of the wavelength coincidence of \ion{O}{III} 303.799~{\AA} and 303.693~{\AA}, be absorbed by \ion{O}{III}. If the latter happens, then the later decay of the excited \ion{O}{III} can produce a cascade of emission lines escaping the region (e.g. 3047, 3133, 3312, 3341, 3444, and 3760 {\AA}\footnote{Unfortunately, our spectra do not cover the 3000-4000~{\AA} range to detect the other \ion{O}{III} Bowen lines.}), and eventually a FUV \ion{O}{III} 374.436~{\AA} photon. The 374.436~{\AA} can then be absorbed by ground-state \ion{N}{III}, which further triggers a cascade of emission lines (\ion{N}{III} 4100, 4640~{\AA}). Bowen fluorescence typically requires a high flux of FUV/ soft-X-ray photons in order to produce the \ion{He}{II} Ly$\alpha$ photons. 

We measure relative line intensities of $F(\mathrm{\ion{He}{II}})/F(\mathrm{H\, \beta}^{\mathrm{b}})\sim0.57$, $F(\mathrm{\ion{N}{III} \, 4640})/F(\mathrm{\ion{He}{II}})\sim0.65$ and $F(\mathrm{\ion{N}{III} \, 4640})/F(\mathrm{H\, \beta}^{\mathrm{b}})\sim0.37$. \citet{Netzer1985} predicted the relative Bowen line intensities in AGNs under a range of different metal gas densities and abundances, where they found that to produce the high $F(\mathrm{\ion{He}{II}})/F(\mathrm{H\, \beta}^b)$ ratios seen in AT~2019avd as well as the high observed $F(\mathrm{\ion{N}{III} \, 4640})/F(\mathrm{H\, \beta}^b)$ ratio, the gas producing the Bowen fluorescence must have very high density ($n_{\mathrm{H}}>10^{9.5}$cm$^{-3}$) and high N and O abundances relative to cosmic abundances.

\subsubsection{Coronal lines}
From the line fitting seen on the WiFeS spectrum in Fig.~\ref{fig:coronal_line_fits}, we infer the luminosities of the [\ion{Fe}{X}]~6375~{\AA} and [\ion{Fe}{XIV}]~5303~{\AA} emission lines to be $\sim 2\times 10^{39}$ and $\sim 3\times 10^{39}$~$\mathrm{erg \, s^{-1}}$. We also infer relative intensities of $F(\mathrm{[\ion{Fe}{X}]\, 6375})/F(\mathrm{[\ion{O}{III}] \, 5007})\sim 2.4$ and $F(\mathrm{[\ion{Fe}{XIV}]\, 6375})/F(\mathrm{[\ion{O}{III}] \, 5007})\sim 3$. Based on the coronal line ratio definitions proposed in \citet{Wang2012}, AT~2019avd is classified as an extreme coronal line emitter (ECLE), where extreme is defined relative to the line ratios seen in coronal line AGNs (e.g. \citealt{Nagao2000} report a maximum line ratio for $F(\mathrm{[Fe\, X]\, 6375})/F(\mathrm{[\ion{O}{III}] \, 5007})$ of 0.24 over a sample of 124 Seyferts). Also, given the non-detected set of [\ion{Fe}{VII}] emission lines in AT~2019avd which are seen in some ECLEs, and relatively weak [\ion{O}{III}]~5007~{\AA} emission, AT~2019avd belongs to the subset of ECLEs that were designated as TDEs in \citet{Wang2012}.

The Fe coronal lines are narrower relative to the \ion{He}{II} and \ion{N}{III} 4640~{\AA} emission lines (Table~\ref{tab:optSpecFWHM_wifes}), with FWHM for the [\ion{Fe}{XIV}]~5303~{\AA} and [\ion{Fe}{X}]~6375~{\AA} of $1560\pm 140$ and $770 \pm 40$~km~s$^{-1}$ respectively. Under the assumption that the line widths are set by the virial motion of the gas, this suggests that the coronal lines are produced further away from the BH than the Bowen lines, and also with the higher ionisation coronal lines being produced closer to the BH than the lower ionisation lines. The width of [\ion{Fe}{XIV}]~5303~{\AA} is comparable to the observed Balmer emission. We also note the differing line profiles of the [\ion{Fe}{XIV}]~5303~{\AA} and [\ion{Fe}{X}]~6375~{\AA} emission, with the latter showing a stronger blue asymmetry (Fig.~\ref{fig:coronal_line_fits}). 

As discussed in \citet{Wang2012}, the weakness of [\ion{Fe}{VII}] emission relative to [\ion{Fe}{X}] and [\ion{Fe}{XIV}] may be explained through the coronal line gas either being overionised under a high X-ray flux, or due to collisional de-excitation of [\ion{Fe}{VII}], because it has a lower critical density ($\sim 10^7\, \mathrm{cm}^{-3}$) compared with the higher ionisation lines ($\sim 10^{10}\, \mathrm{cm}^{-3}$, \citealt{Korista1989}). 

\subsubsection{Black hole mass estimate}\label{sec:mbh_estimation}
We assume that the gas that produces the broad H$\beta$ emission is virialised around the SMBH at the centre of the galaxy, and use the `single epoch' mass-estimation technique (e.g. \citealt{Vestergaard2006}) to infer the black hole mass using the following scaling relation from \citet{Assef2011}:
\begin{equation}
    \log \left( \frac{M_{\mathrm{BH}}}{M_{\odot}} \right) = A + B\log \left( \frac{\lambda L_{\mathrm{\lambda}}}{10^{44} \mathrm{\, erg \, s^{-1}}}\right) + C\log \left( \frac{\mathrm{FWHM}}{\mathrm{km\, s}^{-1}}\right)
,\end{equation}
with $A=0.895$, $B=0.52$ and $C=2$. From the measured FWHM of the broad $H\beta$ component 1420~km~s$^{-1}$ and $L_{5100}=\lambda L_{\lambda}(5000 \AA)\sim 2\times 10^{42}$ erg~s$^{-1}$ from the WiFeS spectrum\footnote{$L_{\lambda}(5100\AA)$ is computed from the mean of $L_{\lambda}$ between 5095 and 5105~{\AA}}, we
then infer $\log [M_{\mathrm{BH}}/M_{\odot}] \sim 6.3$, albeit with a large uncertainty of $\sim$0.3~dex \citep{Assef2011}. We note that using this technique requires the correlations between continuum luminosity and radius of the broad line region (BLR; e.g. \citealt{Kaspi2005}) obtained in previous AGN reverberation mapping experiments to also hold for the BLR around the SMBH in AT~2019avd.

\subsubsection{Baldwin, Phillips, and Terlevich line diagnostic}\label{sec:bpt}
From the fitting of the WiFeS spectrum, we infer line flux ratios of log[[N II]~6583/H$\alpha ^{\mathrm{n}} ]=-0.099^{+0.015}_{-0.016}$ and log[[\ion{O}{III}] 5007/H$\beta ^{\mathrm{n}}]=0.09^{+0.08}_{-0.10}$. According to a Baldwin, Phillips, and Terlevich (BPT) line diagnostic test \citep{Baldwin1981}, such line ratios suggest that a blend of star formation and AGN activity is responsible for producing the narrow line emission in the host galaxy of AT~2019avd \citep{Kauffmann2003,Kewley2006}. Without an archival spectrum though, it is unclear whether the [\ion{O}{III}] 5007~{\AA} and [\ion{N}{II}]~6583~{\AA} lines have increased in intensity since the initial ZTF outburst, or an AGN-like ionising source has always been present. 

\subsection{Mapping out the BLR}
\begin{figure}
    \centering
    \includegraphics[scale=0.8]{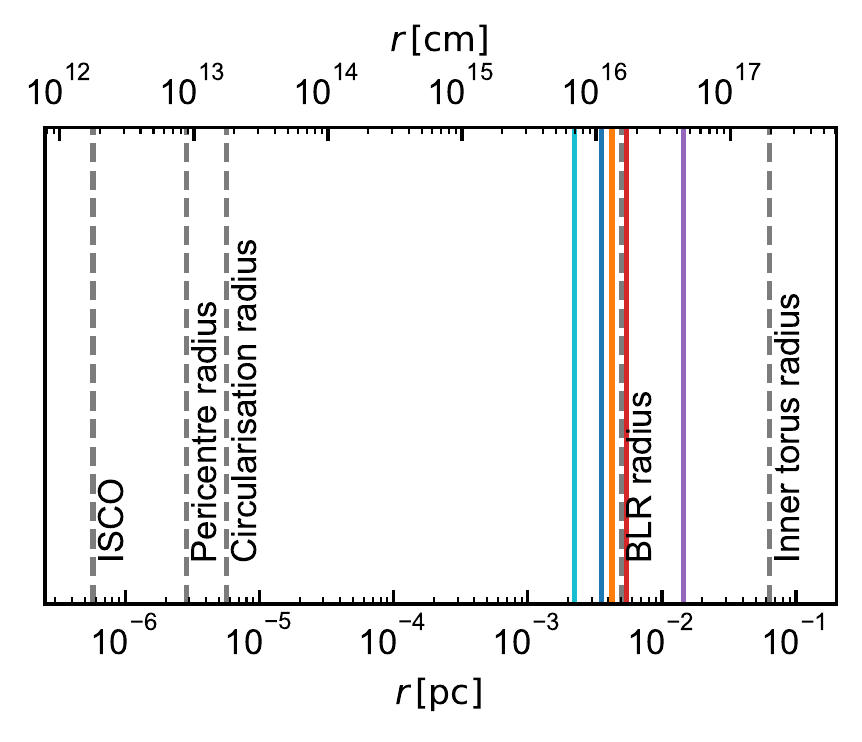}
    \includegraphics[scale=0.8]{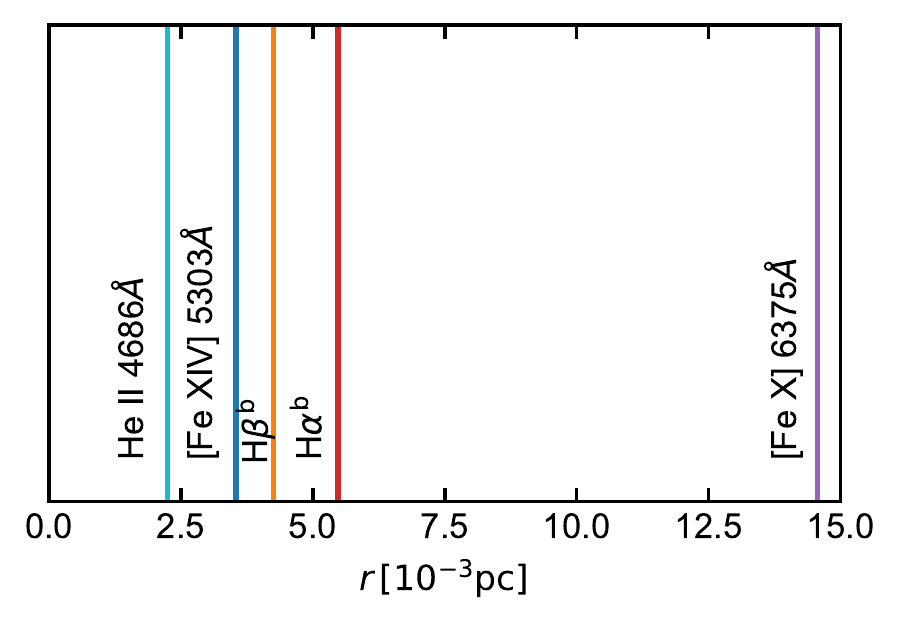}
    \caption{Estimated radii from the BH where different observed optical emission lines are produced in AT~2019avd, compared with various key physical length scales predicted in the literature (assuming $\log [M_{\mathrm{BH}}/M_{\odot}]=6.3$). The pericentre and circularisation radii are computed assuming a Sun-like star incident on this BH with its closest approach at the tidal radius. Similarly to \citet{Kollatschny2014}, we see evidence for a stratified BLR. The coloured lines represent length scales that were obtained based on observations of AT~2019avd, whilst the grey dashed lines are based on various scaling relations in the literature (BLR radius based on \citealt{Kaspi2005}, whilst the inner torus radius was computed using equation 1 of \citealt{Nenkova2008}, assuming a dust sublimation temperature of 1500K).}
    \label{fig:shell_structure}
\end{figure}

Assuming that each observed emission line is broadened due to its virial motion around the central BH, we can use the measured
FWHMs to obtain rough estimates of the distances from the central ionising source at which each line is produced  (Fig.~\ref{fig:shell_structure}). Similar to previous work (e.g. \citealt{Korista1995,Kollatschny2003,Bentz2010}), we also find evidence for a stratified BLR, whereby the higher ionisation lines are produced in regions closer to the BH. 

\section{Discussion}\label{sec:discussion}
Based purely on its X-ray luminosity evolution, AT~2019avd most likely involves an accreting SMBH at the centre of a galaxy. Whilst the large amplitude X-ray flaring (factor of $\gtrsim $600), soft X-ray spectrum, lack of previous strong (and sustained) AGN activity, and the implied unabsorbed X-ray peak luminosity in the $0.2-2$~keV energy range of $2\times10^{43}$ erg~s$^{-1}$ (using spectroscopic $z=0.029$, see section~\ref{sec:opt_spectroscopy}) initially made the source a strong TDE candidate, this is clearly discordant with the double-peaked optical variability seen in the ZTF observations (it does not look like a prototypical, single-event TDE as observed elsewhere). In the following section, we discuss potential origins of the rich phenomenology seen in AT~2019avd.

\subsection{AT~2019avd as non-TDE-induced AGN variability}\label{sec:non_tde}
If AT~2019avd is related to AGN activity that was not induced by a TDE (herein referred to simply as AGN `activity' or `variability'\footnote{As a TDE may transform a quiescent BH into an AGN, the variability in BHs induced by TDEs is also just a subset of AGN variability.}), then the combination of its X-ray and optical light curves make it one of the most extreme cases of AGN variability observed to date. 

It is clear that the X-ray spectrum of AT~2019avd (section~\ref{sec:x-ray_spec_fitting}) is far softer than what is commonly seen in Seyfert 1s; for example, the power-law slope for \textit{Swift} OBSID 00013495001 was $5.3^{+0.4}_{-0.4}$, whilst \citet{Nandra1994} model the observed power-law slope distribution with a Gaussian distribution of mean 1.95 and standard deviation 0.15. However, based on the measured FWHMs of the broad Balmer emission lines in the optical spectrum, it would be classified as a NLSy1, and softer spectral indices have also been observed in the NLSy1 population; a systematic {\it ROSAT} study of this by \citet{Boller1996} found power-law slopes of up to $\sim 5$. NLSy1s are also known to exhibit rapid, large-amplitude X-ray variability (e.g. \citealt{Boller1996}). As the X-ray variability of NLSy1s over longer timescales has not been extensively monitored before,  how common AT~2019avd-like X-ray flares are within this population is currently unclear. For this reason, the X-ray properties alone cannot be used to state that the observed variability in AT~2019avd was induced by a TDE.

\begin{figure}
    \centering
    \includegraphics[scale=0.8]{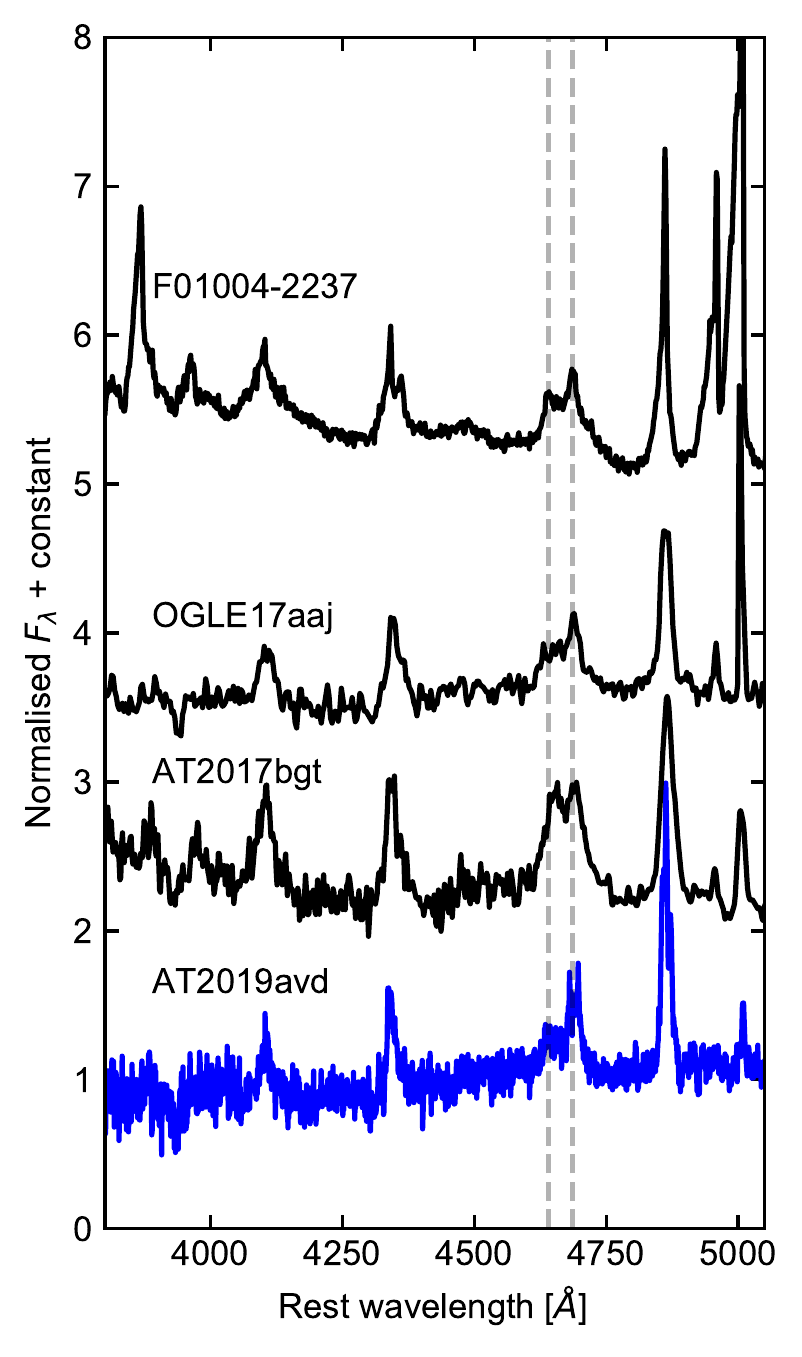}
    \caption{Comparison of the optical spectrum of AT~2019avd with those of the three nuclear transients recently identified as a new class of flares from accreting SMBHs in \citet{Trakhtenbrot2019a}. The two dashed grey lines mark the positions of \ion{N}{III}~4640~{\AA} and \ion{He}{II}~4686~{\AA}. All objects share high $F$(\ion{He}{II} 4686~{\AA})/$F$(H$\beta$), and at least one Bowen emission line (\ion{N}{III}~4640{\AA}).}
    \label{fig:nuclear_transient_comparison_trakh}
\end{figure}

However, AT~2019avd shows a number of features in its optical spectrum that are infrequently seen in NLSy1s. First, NLSy1s commonly show strong \ion{Fe}{II} emission (e.g. \citealt{Rakshit2017}), whereas this is not seen in the WiFeS spectrum, and only a weak \ion{Fe}{II} complex is seen in the NOT spectrum in AT~2019avd. Instead, the most prominent Fe emission we observe are the transient, ECLE-like higher ionisation coronal lines of [\ion{Fe}{XIV}] 5303~{\AA} and [\ion{Fe}{X}] 6375~{\AA} in the WiFeS spectrum. During our spectroscopic follow-up campaign, we also observe the appearance of \ion{He}{II}~4686~{\AA} and \ion{N}{III}~4640~{\AA} emission lines (attributed to Bowen fluorescence). The optical spectrum at late times appears similar to the recently identified new class of flaring transients by \citet{Trakhtenbrot2019a}, and we present a comparison of AT~2019avd with this class in Fig.~\ref{fig:nuclear_transient_comparison_trakh}. Whilst AT~2019avd shares the broad emission feature around 4680{\AA} with the AT~2017bgt flare class, the optical spectrum of AT~2019avd  is distinguishable from the other members based on its much weaker [\ion{O}{III}]~5007{\AA} emission line. A likely reason for this is that the host galaxies of the other flares had persistent, higher luminosity AGNs in them prior to the optical outburst, relative to AT~2019avd. In addition, AT~2019avd's large amplitude, ultra-soft X-ray flare, and its optical light-curve evolution make it unique amongst the AT~2017bgt flare class. 

Finally, we stress that the double-peaked optical variability shown by AT~2019avd is unprecedented for a NLSy1, which when combined with its X-ray properties, make AT~2019avd clearly unique relative to all previous examples of AGN variability. Further examples of NLSy1 variability seen during the ZTF survey will be presented in a separate publication \citep{frederick2020family}.

\subsection{An origin related to tidal disruption?}
\subsubsection{Canonical tidal disruption event}
As AT~2019avd shows a very-large-amplitude, soft-X-ray flare from the nucleus of a galaxy that shows no strong signs of prior AGN activity, it appears similar to the predicted observational signatures for TDEs (e.g. \citealt{Rees1988}) and most of the previous X-ray-selected thermal TDE candidates \citep{Bade1996,Komossa1999a,Komossa1999,Grupe1999,Greiner2000,Saxton2019TidalFuture}. On the other hand, its optical spectrum shows a far weaker blue continuum component relative to that seen in optically selected TDEs, as well as narrower Balmer emission lines (for TDEs where these are detected); based on these two pieces of evidence, it would be straightforward to declare that AT~2019avd is not a TDE candidate, according to criteria for optical TDE selection in \citet{van2020seventeen}.

The observed broad Balmer emission lines in AT~2019avd instead appear more like those commonly seen in the broad emission lines of Seyfert 1s. With such similarity, a mechanism analogous to the broad line emission in AGNs is likely operating in AT~2019avd, whereby the line widths of hydrogen recombination lines are set by the gas kinematics (whereas some TDEs may have line widths set by repeated non-coherent electron scattering; e.g. \citealt{Roth2018}), and the high densities in the BLR result in the line intensity responding effectively instantaneously to changes in the continuum flux. In the limit of a weak TDE-like reprocessing layer\footnote{And likely a lack of optically-selected observed TDE features.}, the optical spectrum of a TDE may appear similar to that of an AGN, as has been previously suggested (e.g. \citealt{Gaskell2014}). The timescales for the evolution of the spectral features in such systems may be different from those observed in optically selected TDEs, as they originate from a region further away from the BH than the reprocessing layer. 

The optical emission mechanism in TDEs is currently not well understood, although it is thought to arise either from shocks produced from stellar debris stream self-intersections \citep{Shiokawa2015a,Piran2015}, or from debris reprocessing the emission from an accretion disc (e.g. \citealt{Loeb1997,Ulmer1998,Roth2016,Roth2018}). However, it is unclear how luminous the shocks are from stream self-intersections, whilst for the reprocessing scenario we still do not understand where the reprocessor is situated, where it forms, how large its covering angle would be from the BH, how efficiently it converts disc emission into the optical wavebands, or how all of these aspects are affected by the properties of the  BH and those of the disrupted star. There is currently not a large enough sample of TDEs selected through \textit{both} X-ray and optical surveys to test these various models of optical emission, and to properly assess the various complex underlying selection effects likely present in the existing TDE candidate population. A key example of these effects is the fact that only a small fraction of optically selected TDEs show transient X-ray emission ($\sim 25$\% of optically selected TDEs in \citealt{van2020seventeen} were X-ray bright); \citet{Dai2018} suggested that the observed properties of a TDE may be dependent upon the viewing angle to the newly formed disc. 

Given the above, and that there are also no X-ray selected, non-relativistic TDEs in the literature that have high-cadence optical photometric light curves available\footnote{Although the 4 X-ray bright TDEs in \citet{van2020seventeen} were monitored at a high cadence with ZTF and \textit{Swift} UVOT, these were optically-selected TDEs.}, we cannot rule out a TDE-related origin for AT~2019avd simply on the basis of a lack of optically selected TDE features in the optical spectrum. However, we do disfavour the canonical TDE interpretation (seen in optically selected TDEs) for this flare on the basis of the double-peaked optical light curve, which has not been observed in any of the TDEs identified by ZTF so far. Secondary maxima have previously been seen in the light curves of some TDE candidates (a compilation is presented in Fig.~8 of \citealt{Wevers2019a}), though not at optical wavelengths and of far smaller amplitude increase compared with AT~2019avd (with the exception of the TDE in an AGN candidate in \citealt{Merloni2015}). 

\subsubsection{A more exotic variant of  a tidal disruption event?}
A large fraction of stars may exist in binary systems (e.g. \citealt{Lada2006}). \citet{Mandel2015} studied the various outcomes of a binary star passing close to a SMBH from a nearly radial orbit. In $\sim 20$\% of such approaches, a double tidal disruption event (DTDE) is produced, whereby both stars in the binary are disrupted in succession. These latter authors estimated that $\sim$10\% of all stellar tidal disruptions could be associated with DTDEs, with such events expected to produce double-peaked light curves. 

We can use the inferred rise-to-peak timescales from the ZTF light curves to test the feasibility of whether AT~2019avd may have been triggered by a DTDE, specifically for the case where each peak is associated with the rise to peak mass fallback of each successive disruption. \citet{Guillochon2013} present the time taken for a single TDE to reach peak mass fallback rate (in their equation A2):
\begin{equation}\label{eqn:t_peak}
    t_{\mathrm{peak}} = B_{\gamma} \left( \frac{M_{\mathrm{BH}}}{10^6M_{\odot}}\right) ^{1/2} \left( \frac{M_{\star}}{M_{\odot}}\right) ^{-1} \left( \frac{R_{\star}}{R_{\odot}}\right) ^{3/2} \, \mathrm{years}
,\end{equation}
where $B_{\gamma}$ is a function of $\beta$, the ratio of the tidal radius of the  BH to the pericentre of the orbit of the star, $\gamma$ is the polytropic index of the star\footnote{We use $\gamma = 4/3$ for $0.3 M_{\odot} < M_{\star} < 22 M_{\odot}$, and $\gamma = 5/3$ for $M_{\star}$ outside this range, as in \citealt{Mockler2019}.}, $M_{\mathrm{BH}}$ is the black hole mass, and $M_{\star}$ and $R_{\star}$ are the mass and radius of the star being disrupted.

Similarly to \citet{Merloni2015}, we then generate a grid of $M_{\star}$ and $\beta$, log-uniformly between (0.1$M_{\odot}$, 100$M_{\odot}$) and (0.5, 4), respectively, and compute $R_{\star}$ for each $M_{\star}$ using the mass--radius relationship for zero-age main sequence stars presented in \citet{Tout1996}. For each possible combination of $M_{\star}$ and $\beta$, and for a black hole with $\log [M_{\mathrm{BH}}/M_{\odot}] \sim 6.3$, we check whether it can produce $t_{\mathrm{peak}}$ (using equation~\ref{eqn:t_peak}) within 20\% of the observed peak timescales in the ZTF light curves ($\sim 24$ days and $\sim 260$ days for the first and second peak respectively). We also enforce the constraint that its tidal radius lies outside of the Schwarzschild radius for the system, so that it can produce a TDE with the star being swallowed whole by the black hole.
\begin{figure}
    \centering
    \includegraphics[scale=0.8]{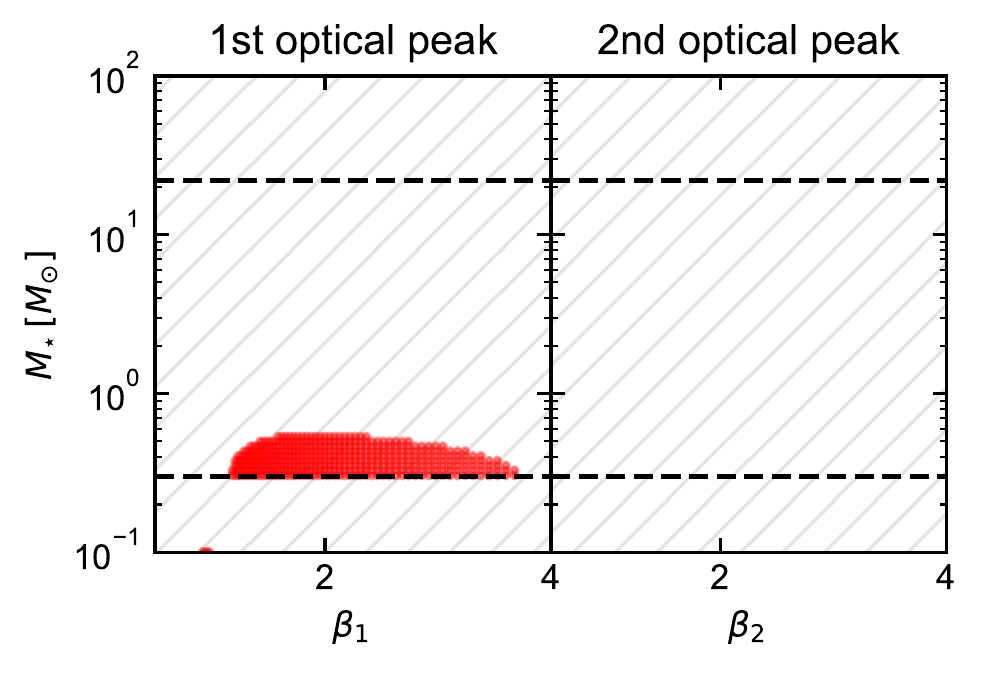}
    \caption{Constraints on the $M_{\star}$, $\beta$ parameter space, obtained for explaining the origin of AT~2019avd as a DTDE on SMBH. Red markers represent a permitted $M_{\star}$, $\beta$ configuration, whilst a region that contains grey hashing represents a configuration that is not able to reproduce the observed timescales for the given peak. Results were obtained for a black hole with $\log [M_{\mathrm{BH}}/M_{\odot}] \sim 6.3$. Since there are no red markers on the second optical peak plot, there is no permitted $M_{\star}$, $\beta$ pairing that can reproduce the observed peak timescale for the second optical peak. The black dashed lines bound $0.3 M_{\odot} < M_{\star} < 22 M_{\odot}$, where we adopt $\gamma = 4/3$.}
    \label{fig:dtde_single_smbh_param_constraints}
\end{figure}
 
We plot the permitted regions of the $M_{\star}$, $\beta$ parameter space in red in Fig.~\ref{fig:dtde_single_smbh_param_constraints}, where we see that no main sequence binary star configuration can reproduce the observed rise times for both the first and second peaks. It would also be possible to obtain further constraints on the feasibility of this scenario based on the observed peak luminosities (similar to \citealt{Merloni2015}) and their ratio, as well as from the inferred properties of the binary itself, such as from the time between the two observed peaks (which could be used to constrain the semi-major axis) and the inferred mass ratio. However, the constraints provided from $t_{\mathrm{peak}}$ are perhaps the simplest to implement and are sufficient to highlight the caveats of a \textit{simple} DTDE interpretation. 

\citet{Bonnerot2019} recently suggested that following the disruption of a stellar binary, the two separate debris streams may collide prior to their fallback onto the black hole. These collisions then shock-heat the gas, and were predicted to produce an optical flare prior to the main flare of the disruption event. Such a model for a binary TDE could potentially explain the observed double-peak light curve, and the observed emergence of the Bowen feature after the second peak (the soft X-rays can only be emitted once the accretion disc has formed). However, a caveat to this interpretation is that both a strong ionising flux and high gas densities are required for Bowen fluorescence to be produced, and we cannot confidently state here that the reason for not observing Bowen lines in the NOT spectrum is the absence of an X-ray-emitting accretion disc during that observation, because the absence of Bowen lines may also be due to insufficiently high gas densities (not all TDEs that are X-ray bright have displayed Bowen emission lines). We do not rule out this {more complex} DTDE scenario for AT~2019avd here, but do not perform a detailed comparison between the simulations in \citet{Bonnerot2019} and AT~2019avd in the present paper. Another alternative could be that AT~2019avd involved some type of TDE about a SMBH binary (e.g. \citealt{Liu2009,Coughlin2017}), where in such systems, the presence of the secondary BH can perturb the accretion flow onto the primary, leading to intermittent light curves. 

\subsection{Could AT~2019avd be supernova-related?}
The spectra of Type IIn SNe can appear similar to those of AGNs (e.g. \citealt{Filippenko1989}), as they can show broad and narrow emission lines, an absence of P-Cygni profiles, and higher luminosities and slower decay timescales relative to normal Type II SNe \citep{Nyholm2020}. Type IIn SNe typically also show the highest X-ray luminosities amongst all SNe. However, AT~2019avd has a $L_{\mathrm{0.2-2keV}}$ that is about an order of magnitude higher than what is seen in most X-ray-luminous Type IIn SNe, when considering the sample of IIn shown in Fig. 3 of \citet{Dwarkadas2012}. Furthermore, the X-ray emission from Type IIn SNe is predicted to be hard (e.g. \citealt{Ofek2013}), whilst that of  AT~2019avd is ultra-soft. Based on the X-ray emission alone, we disfavour the idea that both optical peaks in AT~2019avd are related to a single Type IIn supernova.

Given the observed peak and decay timescales (Fig.~\ref{fig:optical_lc_fits}), the peak absolute magnitude of the optical light curve ($\sim -18.5$), the small amount of reddening seen in the ZTF light curve during the decay phase, and the NOT spectrum, the first optical peak may have been associated with a Type IIn SN. The second optical peak would then be associated with a `turn on' event in the SMBH that sees a vast increase in the accretion rate and the luminosity of the BH. This scenario would then explain why the \ion{He}{II}, Bowen, and coronal lines are not seen in the NOT spectrum, and only in the spectra taken after the second peak. However, the probability of observing both a Type IIn SN and an AGN `turn on' event within just over a year of each other is extremely small given the apparent rarity of extreme `turn-on' events in AGNs (especially those showing an AT~2019avd-like X-ray outburst) and the expected detection rates for Type IIn SNe (e.g. \citealt{Feindt2019}), and we therefore disfavour a scenario where AT~2019avd is the \textit{chance} coincidence of a Type IIn SN and extreme AGN ignition event within roughly one year of each other.

 \section{Conclusions}\label{sec:conclusions}
 This paper presents an overview of a set of multi-wavelength observations of an exceptional nuclear transient, AT~2019avd, whose main observed features are as follows:
\begin{enumerate}
    \item eROSITA detected an ultra-soft ($kT \sim 85$~eV) X-ray brightening ($\gtrsim 90$ times brighter than a previous 3$\sigma$ upper flux limit) from a previously X-ray-inactive galaxy (Section~\ref{sec:xray_observations}). 
    \item AT~2019avd was initially observed on a weekly basis with {\it Swift} XRT/UVOT for 6 weeks following the eROSITA detection. The host had brightened in all UVOT bands by $\sim 1$~mag relative to archival {\it GALEX} observations, and was observed with $0.2-2$~keV X-ray flux consistent with the eROSITA detection (Section~\ref{sec:xray_observations}). A further \textit{Swift} observation $\sim$5 months after the initial eROSITA detection revealed a brightening by a factor of approximately six in the $0.2-2$~keV band relative to the eROSITA detection. AT~2019avd therefore shows a net brightening in the $0.2-2$~keV band by a factor of \textit{at least} 600 relative to the 3$\sigma$ upper detection limit derived from an \textit{XMM-Newton} pointing in 2015.
    \item In the 450 days prior to the eROSITA detection, ZTF observed a double-peaked light curve (Section~\ref{sec:photo_evolution_host_galaxy}). The first optical peak shows rise and decay timescales akin to TDEs and SNe, whilst the rise time of the second peak is more similar to those seen in AGNs. No optical outbursts were detected during ASAS-SN observations over the seven years preceding the initial outburst seen by ZTF.
    \item Optical spectroscopic follow-up finds transient \ion{He}{II} emission, Bowen fluorescence lines, and high-ionisation coronal lines ([\ion{Fe}{X}] 6375~{\AA}, [\ion{Fe}{XIV}] 5303~{\AA}) in the spectra taken after the second optical peak, but not in the spectrum taken 30 days after the first peak. The presence of such a set of lines requires an intense source of soft X-ray emission and extremely high densities. Broad Balmer emission lines were detected in spectra 30 days after the first peak in the ZTF light curve, as well as in all spectra taken in the weeks after the eROSITA detection with FWHM~$\sim 1400$km~s$^{-1}$ (Section~\ref{sec:opspec_data_analysis}).
\end{enumerate}
 
AT~2019avd thus shows a set of observed features which have never been observed together in the same nuclear transient before, and further complicates the non-trivial task of distinguishing the physical origin of large-amplitude variability seen in galactic nuclei. Whilst a discussion on the potential origins of this transient is presented in Section~\ref{sec:discussion}, it is still unclear what has triggered such exotic behaviour. Detailed simulations would be welcome to distinguish between the various possible scenarios. These will be well complimented with future planned observations (\textit{Swift}, \textit{NICER}, \textit{XMM-Newton}) monitoring the late-time evolution of AT~2019avd. Finally, we note that during its eight successive all-sky surveys in the following years, eROSITA will systematically monitor the X-ray variability of AGNs and map out the population of nuclear transients. With this information, we will be able to better understand the extent of the X-ray variability shown by AT~2019avd, and make a more informed judgement on the origin of this transient.

%-----------------------------------------------------------------

\begin{acknowledgements}
We thank the anonymous referee, and the journal editor, Sergio Campana, for constructive comments which helped improve this paper. AM thanks the Yukawa Institute for Theoretical Physics 
at Kyoto University, where discussions held during the YITP workshop 
YITP-T-19-07 on International Molecule-type Workshop 
"Tidal Disruption Events: General Relativistic Transients” 
were useful to complete this work. AM thanks Mariuz Gromadzki, Giorgos Leloudas and Clive Tadhunter for sharing optical spectra. A.M. acknowledges support from and participation in the International Max-Planck Research School (IMPRS) on Astrophysics at the Ludwig-Maximilians University of Munich (LMU). 

BJS is supported by NSF grant AST-1907570. BJS is also supported by NASA grant 80NSSC19K1717 and NSF grants AST-1920392 and AST-1911074. 

BT acknowledges support from the Israel Science Foundation (grant number 1849/19)

IA is a CIFAR Azrieli Global Scholar in the Gravity and the Extreme Universe Program and acknowledges support from that program, from the European Research Council (ERC) under the European Union’s Horizon 2020 research and innovation program (grant agreement number 852097), from the Israel Science Foundation (grant numbers 2108/18 and 2752/19), from the United States - Israel Binational Science Foundation (BSF), and from the Israeli Council for Higher Education Alon Fellowship.

L.T. acknowledges support from MIUR (PRIN 2017 grant 20179ZF5KS).

GEA is the recipient of an Australian Research Council Discovery Early Career Researcher Award (project number DE180100346), funded by the Australian Government.

This research was partially supported by the Australian Government through the Australian Research Council's Discovery Projects funding scheme (project DP200102471).

This work is based on data from eROSITA, the primary instrument aboard SRG, a joint Russian-German science mission supported by the Russian Space Agency (Roskosmos), in the interests of the Russian Academy of Sciences represented by its Space Research Institute (IKI), and the Deutsches Zentrum für Luft- und Raumfahrt (DLR). The SRG spacecraft was built by Lavochkin Association (NPOL) and its subcontractors, and is operated by NPOL with support from the Max Planck Institute for Extraterrestrial Physics (MPE).

The development and construction of the eROSITA X-ray instrument was led by MPE, with contributions from the Dr. Karl Remeis Observatory Bamberg, the University of Hamburg Observatory, the Leibniz Institute for Astrophysics Potsdam (AIP), and the Institute for Astronomy and Astrophysics of the University of Tübingen, with the support of DLR and the Max Planck Society. The Argelander Institute for Astronomy of the University of Bonn and the Ludwig Maximilians Universität Munich also participated in the science preparation for eROSITA.

This work was based on observations obtained with the Samuel Oschin Telescope 48-inch and the 60-inch Telescope at the Palomar Observatory as part of the Zwicky Transient Facility project. ZTF is supported by the National Science Foundation under Grant No. AST-1440341 and a collaboration  including Caltech, IPAC, the Weizmann Institute for Science, the Oskar Klein Center at Stockholm University, the University of Maryland, the University of Washington, Deutsches Elektronen-Synchrotron and Humboldt University, Los Alamos National Laboratories, the TANGO Consortium of Taiwan, the University of Wisconsin at Milwaukee, and Lawrence Berkeley National Laboratories. Operations are conducted by COO, IPAC, and UW.  SED Machine is based upon work supported by the National Science Foundation under Grant No. 1106171.

This publication makes use of data products from the Near-Earth Object Wide-field Infrared Survey Explorer (NEOWISE), which is a joint project of the Jet Propulsion Laboratory/California Institute of Technology and the University of Arizona. NEOWISE is funded by the National Aeronautics and Space Administration.

We thank the Las Cumbres Observatory and its staff for its continuing support of the ASAS-SN project. ASAS-SN is supported by the Gordon and Betty Moore Foundation through grant GBMF5490 to the Ohio State University, and NSF grants AST-1515927 and AST-1908570. Development of ASAS-SN has been supported by NSF grant AST-0908816, the Mt. Cuba Astronomical Foundation, the Center for Cosmology and AstroParticle Physics at the Ohio State University, the Chinese Academy of Sciences South America Center for Astronomy (CAS- SACA), and the Villum Foundation. 

The Pan-STARRS1 Surveys (PS1) have been made possible through contributions of the Institute for Astronomy, the University of Hawaii, the Pan-STARRS Project Office, the Max-Planck Society and its participating institutes, the Max Planck Institute for Astronomy, Heidelberg and the Max Planck Institute for Extraterrestrial Physics, Garching, The Johns Hopkins University, Durham University, the University of Edinburgh, Queen's University Belfast, the Harvard-Smithsonian Center for Astrophysics, the Las Cumbres Observatory Global Telescope Network Incorporated, the National Central University of Taiwan, the Space Telescope Science Institute, the National Aeronautics and Space Administration under Grant No. NNX08AR22G issued through the Planetary Science Division of the NASA Science Mission Directorate, the National Science Foundation under Grant No. AST-1238877, the University of Maryland, and Eotvos Lorand University (ELTE).

This work was partially based on observations made with the Nordic Optical Telescope, operated at the Observatorio del Roque de los Muchachos, La Palma, Spain, of the Instituto de Astrofisica de Canarias. Some of the data presented here were obtained with ALFOSC, which is provided by the Instituto de Astrofisica de Andalucia (IAA) under a joint agreement with the University of Copenhagen and NOTSA.
\end{acknowledgements}

   \bibliographystyle{aa} % style aa.bst
   \bibliography{at2019avd} % your references Yourfile.bib

\begin{appendix} %First appendix
\section{Optical spectrum and light-curve fitting}
In Table~\ref{tab:lightcurve_priors}, we list the priors adopted in the fitting of the ZTF/ SEDM light curves, whilst in Table~\ref{tab:op_spec_high_res_fit_components}, we list the priors used in our fitting of the NOT and WiFeS optical spectra.
\begin{table}
        \centering
        \caption{Priors adopted in the fitting of the ZTF light curves. The rise and decay timescales are in units of days, whilst $t_{\mathrm{peak}}$ is in MJD. $F_{\mathrm{max}}$ refers to the maximum observed flux within the given peak.}  
        \label{tab:lightcurve_priors}
        \begin{tabular}{lccr} 
                \hline
                  & Priors \\
                \hline
                Peak 1 & \vtop{\hbox{\strut $\log [\tau_{r, g}] \sim \mathcal{U}(0, \log [300])$,   $\log [\sigma_{r, g}] \sim \mathcal{U}(0, \log [300])$}\hbox{\strut $\log [F_{\mathrm{peak}, r}] \sim \mathcal{U}(\log [0.9F_{\mathrm{max}, r}], \log[10F_{\mathrm{max}, r}])$} \hbox{\strut $\log [F_{\mathrm{peak}, g}] \sim \mathcal{U}(\log [0.9F_{\mathrm{max}, g}], \log[10F_{\mathrm{max}, g}])$} \hbox{\strut $t_{\mathrm{peak}}\sim \mathcal{U}(58450, 58650)$}} \\
                Peak 2 & \vtop{\hbox{\strut $\log [\tau_{r, g}] \sim \mathcal{U}(0, \log [300])$}   \hbox{\strut $\log [F_{\mathrm{peak}, r}] \sim \mathcal{U}(\log [0.9F_{\mathrm{max}, r}], \log[10F_{\mathrm{max}, r}])$} \hbox{\strut $\log [F_{\mathrm{peak}, g}] \sim \mathcal{U}(\log [0.9F_{\mathrm{max}, g}], \log[10F_{\mathrm{max}, g}])$} \hbox{\strut $t_{\mathrm{peak}}\sim \mathcal{U}(59000, 59300)$}}  \\
                \hline
        \end{tabular}
\end{table}

\begin{table}
        \centering
        \caption{Overview of the varying set of Gaussians used for modelling the emission lines in the NOT and WiFeS spectra.} 
        \label{tab:op_spec_high_res_fit_components}
        \begin{tabular}{p{1.85cm}p{6cm}} 
                \hline
                 Region & Components \\
                \hline
                H$\gamma$ &  Single Gaussian for each of H$\gamma$ and [\ion{O}{III}] 4363~{\AA}.   \\
                \ion{He}{II} &  Single Gaussian component for each of \ion{He}{II} 4686~{\AA}, and [\ion{N}{III}] 4640~{\AA}. \\
                H$\beta$ &  Broad and narrow Gaussian component. \\
                H$\alpha$ &  Broad and narrow Gaussian component for H$\alpha$, single Gaussian for each of [\ion{N}{II}] 6549 and 6583~{\AA}.   \\
                {[}\ion{S}{II}] doublet &  Single Gaussian for each of [\ion{S}{II}] 6716 and 6731~{\AA}.  \\
                {[}\ion{O}{III}] 5007~{\AA}, [\ion{Fe}{XIV}] 5303~{\AA}, [\ion{Fe}{X}] 6375~{\AA} &  Single Gaussian for each.  \\
                \hline
        \end{tabular}
\end{table}

\section{Long-term light curve of AT~2019avd}
\begin{figure*}
    \centering
    \includegraphics[scale=0.8]{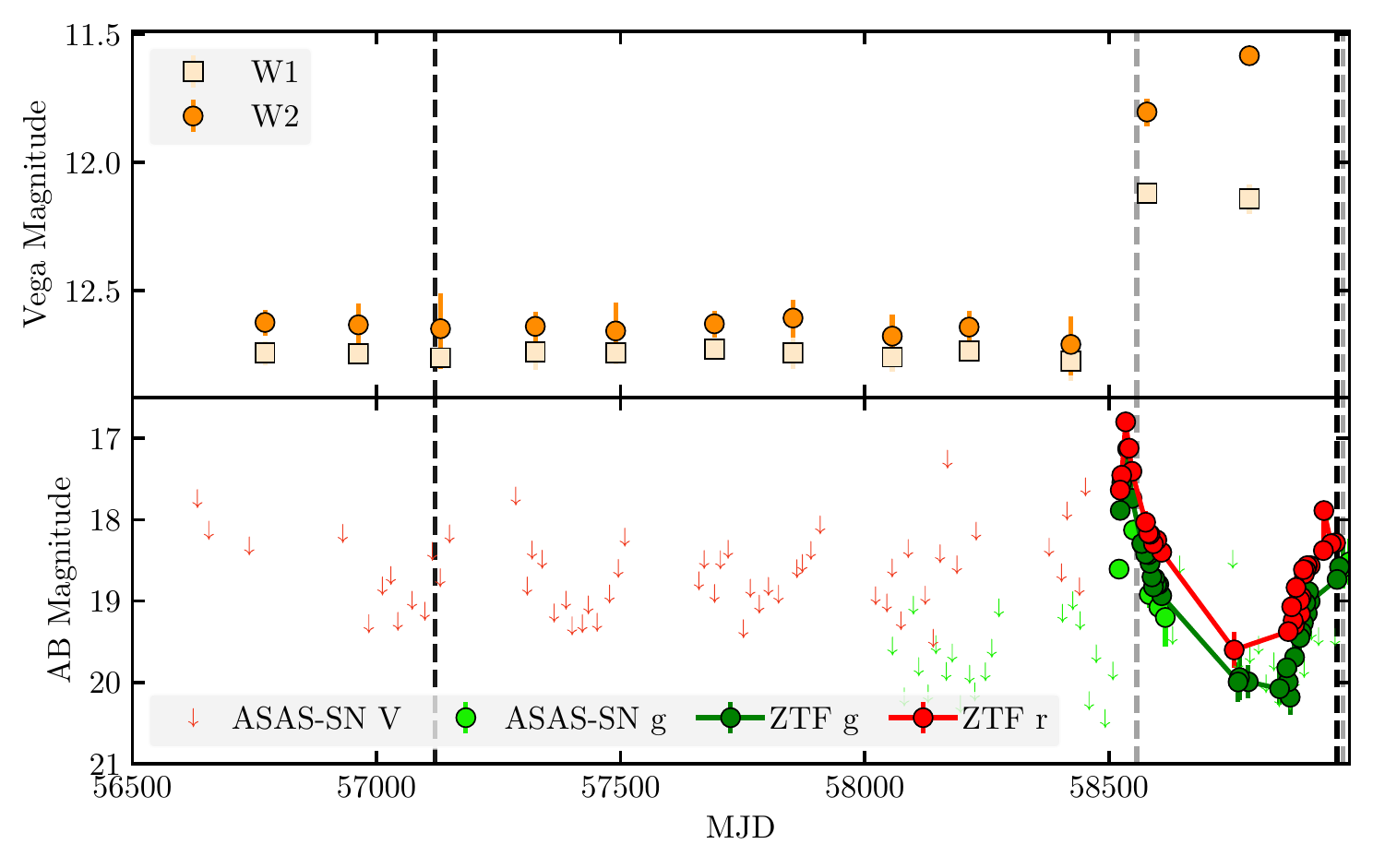}
    \caption{Long-term NEOWISE-R, ASAS-SN, and ZTF light curves of AT~2019avd. The early and late black dashed lines mark the 2015 \textit{XMM-Newton} pointed and the 2020 eROSITA eRASS1 observations respectively. The early and late grey dashed lines mark the MJD that the NOT and first FLOYDS spectra were taken.}
    \label{fig:archival_lightcurve}
\end{figure*}

In Fig.~\ref{fig:archival_lightcurve}, we plot the long-term light curve of AT~2019avd, including the ASAS-SN data. ASAS-SN \citep{Shappee2014} observed the location of AT 2019avd in $V$-band from February 2012 to November 2018 and in $g$-band from October 2017 to September 2020 (the time of writing).  The $V$- and $g$-band observations were reduced using a fully automated pipeline detailed in \citet{Kochanek2017TheV1.0} based on the ISIS image subtraction package \citep{Alard1998,Alard2000}. During each visit, ASAS-SN observed three 90-second dithered images that are then subtracted from a reference image. For the $g$-band we modified the standard pipeline and rebuilt the reference image without any images with $\rm{JD}\geq 2\,458\,518$ to prevent any flux contamination from the outbursts.

All subtractions were inspected manually to remove data with clouds, cirrus, or other issues.  We note, however, that the ASAS-SN light curve was negatively affected by two factors.  First, there is a bright nearby star that is not resolved from the host galaxy in ASAS-SN data and added noise to the subtractions.  Second, the location of AT 2019avd is right on the edge of two ASAS-SN fields.  To help alleviate these issues and increase the ASAS-SN limiting magnitude we stacked the subtractions within a maximum of 10 days.  We then used the IRAF package \texttt{apphot} to perform aperture photometry with a two-pixel, or approximately $16.\!\!''0$, radius aperture on each subtracted image, generating a differential light curve. The photometry was calibrated using the AAVSO Photometric All-Sky Survey \citep{2015AAS...22533616H}. 
\end{appendix}

\end{document}